# Strong substrate strain effects in multilayered WS$_2$ revealed by high-pressure optical measurements

Robert Oliva,[1,*] Tomasz Wozniak,[1] Paulo Eduardo de Faria Junior,[2] Filip Dybala,[1] Jan Kopaczek,[1] Jaroslav Fabian,[2] Paweł Scharoch,[1] Robert Kudrawiec[1]

[1] Department of Semiconductor Materials Engineering, Faculty of Fundamental Problems of Technology, Wroclaw University of Science and Technology, Wybrzeże Wyspiańskiego 27, 50-370 Wrocław, Poland

[2] Department of Physics, University of Regensburg, 93040 Regensburg, Germany.

[*] Corresponding author: robert.oliva.vidal@pwr.edu.pl

**Abstract**

The optical properties of two-dimensional materials can be effectively tuned by strain induced from a deformable substrate. In the present work we combine first-principles calculations based on density functional theory and the effective Bethe-Salpeter equation with high-pressure optical measurements in order to thoroughly describe the effect of strain and dielectric environment onto the electronic band structure and optical properties of a few-layered transition-metal dichalcogenide. Our results show that WS$_2$ remains fully adhered to the substrate at least up to a −0.6% in-plane compressive strain for a wide range of substrate materials. We provide a useful model to describe effect of strain on the optical gap energy. The corresponding experimentally-determined out-of-plane and in-plane stress gauge factors for WS$_2$ monolayers are −8 and 24 meV/GPa, respectively. The exceptionally large in-plane gauge factor confirm transition metal dichalcogenides as very promising candidates for flexible functionalities. Finally, we discuss the pressure evolution of an optical transition closely-lying to the A exciton for bulk WS$_2$ as well as the direct-to-indirect transition of the monolayer upon compression.

## I. Introduction

Atomically thin layers of transition metal dichalcogenides (TMDCs) emerged as a fascinating family of two-dimensional semiconductors owing to their excellent optical properties resulting from their large excitonic binding energies far exceeding the thermal energy.[1] This feature combined with the particular electronic band structure of TMDCs account for exotic valleytronic and spintronic properties which remain present even in the multilayer and bulk form.[2,3] The ease to exfoliate and create heterostructures not only allow to envision novel optoelectronic and spintronic applications, but also flexible devices.[4,5] In order to fully exploit these promising features in TMDCs it is crucial to understand the effect of strain and substrate on their optical and electronic properties.

So far, intensive research has been focused on the effects of substrate and tensile strain on the optical and electrical properties of 2D materials. Importantly, it has been shown that it is possible to reversibly apply 1% uniaxial tensile strain on TMDCs monolayers for over 20 cycles.[6,7] Moreover, substrate-induced tensile strain up to 2.6 % has been reported without fracturing.[8] At such high values, the band gap can be effectively redshifted by approximately 100 meV. While most of the works have been devoted to the study of particular case of the in-plane uniaxial tensile for monolayers,[9,10,7,11–19] the amount of works studying monolayers and multilayers in the biaxial tensile strain case is significantly reduced and its methodologies are varied, including investigation of bubbles present in the layers,[20–23] indirectly tuning the strain through thermally expanding the substrate,[24–28] electromechanically controlled piezoelectric substrates,[29] textured substrates[30] or mechanically bending the substrate.[31] Owing to intrinsic difficulties in estimate strain values in the different methodologies, dispersive values were reported for the gauge factors of the band gap, spanning from 4 meV/%[28] to 124 meV/%[32] for biaxial strained monolayer $MoS_2$. Moreover, the amount of experimental works studying more generalized strain cases, such as compressive; *i)* hydrostatic, *ii)* in-plane or *iii)* out-of-plane strain is scarce. Previous measurements performed in hydrostatic high-pressure devices proved difficult to quantitatively interpret because the sample's strain state is difficult to assess by optical methods.[33,33–37,38] Hence, in order to understand the strain effect on the optical properties of two-dimensional materials from a general perspective, it is highly desirable to systematically study 2D materials under different and controllable strain conditions.

Pressure is a thermodynamic variable that allows to directly probe the interatomic distances without undesired secondary effects arising from changes in temperature, such as anharmonic effects or the modification of the band electron population. In this regard, high-pressure (HP) optical measurements represent a unique tool not only useful to test state-of-the-art band structure calculation methods on two-dimensional materials, but also to assign optical features in complex excitonic systems such as TMDCs. In particular, HP photoreflectance (HP-PR) spectroscopy is a versatile, inexpensive technique that allows to determine the energy of direct optical transitions in a fast and contactless manner.[39] To date, PR measurements have been applied on a great variety of TMDCs at ambient pressure allowing to determine their direct electronic transitions, and detailed information about the band structure as a function of temperature and layer thickness.[40–42] HP-PR measurements allowed to establish the pressure coefficients of the A and B excitonic transitions in the $MoX_2$ and $WX_2$ system (X=S, Se).[43] Moreover, since HP-PR measurements are sensitive to the bulk states, the presence of hidden spin-polarized bands in bulk and multilayered TMDCs could be experimentally confirmed, thus extending the range of spintronic applications from monolayer to bulk, heterostructures and multilayers.[2,3] Despite these works, the effect of hydrostatic pressure on monolayer and few-layer compounds have proven difficult to evaluate due to strong substrate effects. Indeed, highly dispersed values have been provided in the literature for the pressure coefficient of several TMDCs. For the case of $WS_2$ monolayers on different substrates, these values range from 30 (Ref. [34]) to 540 (Ref. [33]) meV/GPa, which made it difficult to draw conclusions on the strain-dependent optical, vibrational and structural properties of TMDCs. Hence, a systematic experimental study about the effect of substrate and number of layers on the high-pressure dependence of the optical properties combined with robust first-principles calculations is highly desirable in order to better understand the effect of in-plane and out-of-plane strain on the electronic band structure.

Here we provide a thorough study of the effect of hydrostatic pressure on the optical transitions of $WS_2$ samples with varying thicknesses, including monolayer (1L), bilayer (2L), trilayer (3L) and bulk deposited on different substrates by means of HP-PR measurements as well as first-principles calculations based on the density functional theory (DFT) and the effective Bethe-Salpeter equation (BSE). Our measurements allowed to determine the pressure dependence of the excitonic transitions in $WS_2$ as a function of sample thickness, from monolayer to bulk. Owing to the good agreement between experimental and theoretical calculations, our results allowed to shed new light onto the effect of strain on the band gap of TMDCs and gauge factors are provided for $WS_2$. Finally, we propose a useful model to describe the effect of strain in two-dimensional systems for generalized strain conditions, including the impact of a substrate's material on the band gap pressure coefficient in HP measurements.

## II. Methods
### a) Experimental details

A series of WS$_2$ monolayer and multilayer (up to 3 layers) samples were epitaxially grown by CVD and deposited on different substrates with a coverage area exceeding 95%. In order to explore different strain conditions, the monolayers were deposited on both soft and hard commercially available and widely used substrates, from one side quartz (SiO$_2$) and silicon (Si) and from the other sapphire (Al$_2$O$_3$), respectively. All multilayers were deposited only on sapphire substrates. These samples were cut (≈10 mm² area) and mounted inside a high-pressure UNIPRESS piston cylinder cell, where pressure was generated using a mechanical press. The chosen hydrostatic medium for the high pressure measurements was Daphne 7474, which remained hydrostatic and transparent during the whole experiment (up to a pressure of 180 GPa). Since the value of the static dielectric constant of the pressure transmitting medium (PTM) is relevant for the interpretation of measured excitonic transitions at room and high pressure (as elaborated in the Discussion section) we measured, for the first time, the corresponding value of Daphne 7474 using the capacitance method,[44] and found a value of ε = 2.022(1). The impedance was measured with Agilent 4294A precision impedance analyzer.

The pressure inside the UNIPRESS cell was determined by measuring the resistivity of a InSb probe which provides a 0.01 GPa sensitivity. A sapphire window in the cell provided optical access to the sample, and photoreflectance (PR) and photoluminescence (PL) measurements were performed. For the PR measurements we used a single grating of 0.55 m focal length to disperse the light reflected from the samples. The signal was measured using an InGaAs (Si) detector for energies below (above) 1.25 eV. A chopped (270 Hz) 405 nm laser line was pumped into the sample together with a probe tungsten lamp (power of 150 W). Phase-sensitivity detection of the PR signal was processed with a lock-in amplifier. Further details on the experimental setup can be found elsewhere.[45] All measurements were performed at ambient temperature and pressures up to ≈1.70 GPa. All spectra were recorded during the upstroke except for the spectra on WS$_2$/Si monolayer at pressures below 0.8 GPa for better statistics.

### b) DFT calculations details

In order to elucidate the experimental results and provide additional insights into the pressure dependence of the excitonic transitions observed in the optical spectra, we perform a systematic analysis of the electronic band structure using density functional theory and excitonic effects within the effective BSE. Our calculations not only support the measurements reported above, but also suggest what types of structures future experiments could employ to obtain pressure coefficients similar to those of freestanding layers.

DFT calculations were performed within projector augmented wave (PAW) method[46] in Vienna Ab-initio Simulation Package (VASP),[47] employing the Perdew-Burke-Ernzerhof (PBE) parametrization of generalized gradients approximation (GGA) to exchange-correlation functional.[48] A plane-wave basis cutoff of 500 eV and a 12×12×6 (12×12×1) Γ-centered Monkhorst-Pack grid of k points were chosen for bulk (2D) structures. The BZ integrations were conducted using Gaussian smearing of 0.05 eV. Lattice parameters and atomic positions were optimized to obtain residual stress lower than 0.05 GPa and forces lower than 0.01 eV/Å. A semi-empirical D3 correction for vdW interactions was employed.[49] Elastic constants and static dielectric tensor components (with local field effects included) were evaluated within Density Functional Perturbation Theory. Spin-orbit interactions were taken into account during all the calculations.

Geometry optimization under hydrostatic pressure for bulk WS$_2$ was performed by setting the target Pulay stress as an input. For the case of freestanding 1L, we followed the procedure presented in Ref.[50] assuming the thickness of 1L to be a half of the out-of-plane bulk lattice constant. The out-of-plane stress was evaluated from standard relation $\sigma_\perp = F/A$, where F is the force normal to XY plane and A is its surface area in the unit cell. The vertical positions of outermost atoms were set by hand and fixed, while the remaining atoms positions were relaxed. It should be noted that, on the basis of Newton's third law, the force vectors are pointed outside the structure, as they counteract the external hydrostatic pressure.

### c) BSE calculations details

The excitonic effects are incorporated in our calculations by evaluating the exciton binding energies via the effective BSE equation.[51–55] The two important factors required for the BSE are the conduction and valence band that host the exciton as well as the electrostatic potential that mediates the electron-hole interaction. For the bulk case, the in-plane ($k_x$; $k_y$) band dispersion is treated within the effective mass approach around the K-point while the out-of-plane ($k_z$) band dispersion covers the whole length of −H to H of the Brillouin zone, with the K-point at $k_z = 0$ (see Fig. 7(c) for the schematics of the first Brillouin zone). The band dispersion is then written as

$$E(k_x, k_y, k_z) = \frac{\hbar^2}{2m_n^*}(k_x^2 + k_y^2) + f(k_z), \qquad [1]$$

in which $m_n^*$ is the band effective mass and f($k_z$) is a function that covers all the $k_z$ dispersion from −H to H. The electron-hole interaction is treated via the anisotropic Coulomb potential[56,57] since $\varepsilon_{xx} = \varepsilon_{yy} \neq \varepsilon_{zz}$. The in-plane effective masses, the dispersion f($k_z$) and the dielectric constants are taken from the DFT calculations (see Table S-I in SM). For 1L, 2L and 3L, we assumed parabolic band dispersions around the K-point with the electron-hole interaction given by the Rytova-Keldysh potential,[58,59] with exciton reduced masses and polarization lengths calculated by DFT (see Table S-II in SM). The effect of inhomogeneous dielectric environment is included by averaging the dielectric constants of bottom (substrate) and top (PTM) materials.[55] The effective BSE is then solved numerically with the exciton binding energies obtained from a linear extrapolation using different samplings of the k-grid. For bulk, we used a 3D k-grid with in-plane ($k_x$; $k_y$) components taken from $-k_L$ to $k_L$ with $k_L = 0.2$ Å$^{-1}$ and $k_z$ from $-H$ to $H$ points discretized with $(2N_k + 1)^3$ k-points with $N_k = \{18; 19\}$. For the monolayer case, we used a 2D grid from $-k_L$ to $k_L$ sampled with $2N_k + 1$ points in both $k_x$ and $k_y$ directions (with a total of $(2N_k + 1)^2$ k-points) using $k_L = 0.5$ Å$^{-1}$ and $N_k = \{55; 60\}$.

We perform additional calculations of the exciton binding energies in bulk crystal within effective Gerlach–Pollmann model,[60] which has been successfully applied to bulk MoS$_2$ crystal recently[59].

### III. Experimental results

#### a) Bulk WS$_2$

The PR spectra of bulk WS$_2$ is shown in Fig. 1. As it can be seen from the top panel, three optical transitions blueshift with increasing pressure. These correspond to the A and B excitonic transitions, as well as a weak transition 70 meV above the A transition, labeled A*. All spectra have been fitted with the Aspnes formula[61] (dashed lines in Fig.1-a),

$$\frac{\Delta R}{R} = Re\left[\sum_{j=i}^{n} C_j e^{i\theta_j}(E - E_j + i\Gamma_j)^{-m}\right], \qquad [2]$$

where three transitions are considered ($n=3$) and $m=2$ is taken for excitonic transitions. The phase of the resonances $\theta_j$ was taken from the spectra acquired at ambient pressure, while the amplitude of the resonance, $C_j$, the energy of the transitions $E_j$, and broadening parameter, $\Gamma_j$ were left as free parameters to be fitted. The pressure dependence of the fitted transition energies is plotted in Fig. 1-b. The pressure coefficients are extracted by linearly fitting the data. As it can be seen in the figure, the pressure coefficient of the B transition is ≈20% larger than that of the A transition, while the pressure coefficient of the A* is very similar to that of A.

The origin of the A* excitation has been a matter of debate since in general it could be attributed to either *i)* an excited state of the A transition ($n = 2$), *ii)* a transition from the H-point of the Brillouin zone, *iii)* an interlayer exciton or *iv)* charged excitons such as negative trions.[41,62–66] The latter interpretation can be ruled out since signal from trions are typically smeared out at temperatures above 200 K due to thermal exciton-electron scattering effects.[67] Also, for the case of WS$_2$, the interlayer exciton interpretation can be ruled out since the dark exciton energy is lower than that of the bright,[68] while the A* feature seen in Fig. 1 exhibits higher energy than the A transition. Hence the A* signal can be either attributed to an excited state or an H-point exciton.

High-pressure experiments could prove very useful to resolve such ambiguity since the effect of pressure on their energies is opposite between both interpretations. From one hand an excited state (*n*=2) would exhibit a slightly lower pressure coefficient than the A transition due to a reduction of the excitonic binding energy.[69] From the other hand, the transition at the H k-point would exhibit a significantly larger pressure coefficient, around 42.2 meV/GPa, as predicted from DFT calculations (see theoretical calculations section for more details). Here, the measured pressure coefficient of the A*, around 34 meV/GPa, is significantly higher than that of A (around 24 meV/GPa). While our results suggest that the signal of the A* transition is consistent with an exciton at the H k-point, the associated uncertainties with the fitting procedure for such weak feature make it hard to draw solid conclusions. High-pressure experiments reaching higher pressures and/or at low temperatures are highly desirable in order to unambiguously clarify the origin of the A* feature.

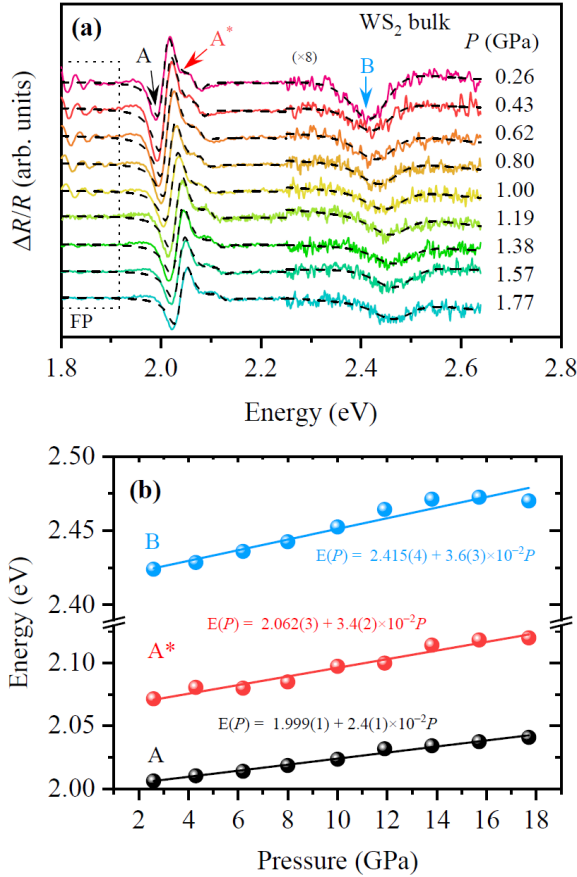

**Fig. 1.**

**(a)** Photoreflectance spectra of $WS_2$ acquired at different pressures. Signal at energies larger than 2.25 eV has been phased-shifted and scaled by a factor of 8 for clarity. Three distinctive PR features can be observed corresponding to the A, A* and B excitonic transitions. The fits to the data are shown as black dashed curves. At energies below the band gap energy, Fabry-Perot interferences show up.

**(b)** Fitted energies of the A, A* and B excitonic transitions as a function of pressure. Linear fits shown as solid lines are used to extract the respective pressure coefficients.

### b) Monolayer $WS_2$ deposited on different substrates

In order to evaluate the impact of the substrate on the pressure coefficient of a $WS_2$ monolayer high-pressure PR experiments are performed on $WS_2$ samples deposited on different substrates. The spectra are shown in Fig. 2 for a samples deposited on; (a) sapphire (Al$_2$O$_3$), (b) silicon (Si) and (c) glass (SiO$_2$). As it can be seen in the figure, single resonances show up around 2.05 eV, which corresponds to the energy of the A excitonic transition. Since the sample deposited on glass exhibited large PL signal, it is included in the panel (c) as well. In order to compare the PL signal to the PR signal, the moduli (Δρ) has been plotted instead, which is obtained from Kramers-Kronig analysis of the complex photoreflectance function,

$$\Delta\rho(E) = \sqrt{\Delta\rho_R^2(E) + \Delta\rho_I^2(E)}, \qquad [3]$$

where the real component is the signal, $\Delta\rho_R(E) = \Delta R/R(E)$, and the imaginary component of the complex PR function is

$$\Delta\rho_I(E) = \frac{2E}{\pi} P \int \frac{\Delta R}{R}(E') \frac{1}{E^2 - E'^2} dE'. \qquad [4]$$

By comparing the PL and moduli of PR signal (Fig. 2-c) we find that the Stokes shift of a WS$_2$ monolayer deposited on glass is 27 meV, much higher than the Stokes shift for our sample on Si, around 3 meV (shown in Fig. S1 of the S.I.). Moreover, the lineshape of the WS$_2$/SiO$_2$ peak exhibits a larger broadening and asymmetry (i.e. the linewidth of the PR signal for the sample on glass, ≈90 meV, is larger than that for the sample on Si, ≈30 meV, or Al$_2$O$_3$, ≈54 meV). Such striking difference in the spectrum may be explained by a larger charge transfer from the substrate to the monolayer. Such doping would result, among other effects, in enhanced signal from charged excitons (trions) and possibly in a stronger exciton-phonon coupling.[70]

While some works investigated the effect of tensile strain on the Stokes shift,[10] it is worth noting that the (compressive hydrostatic) method here employed is limited to relatively low strain values but allowed to measure the direct-to-indirect transition, which takes place below 2 GPa (see Fig. S2 in the S.I.). For the case of our sample grown on Si, we estimate that the transition takes place around 0.72 GPa which is the pressure at which the PL signal vanishes. This value is close to our calculated transition pressure for a freestanding monolayer, as discussed in the theoretical calculations section. The direct-to-indirect transition pressure takes place at a lower pressure for the sample deposited on glass (0.53 GPa) as a consequence of increased in-plane strain due to an increased compressibility of the glass substrate as compared to Si. This substrate-induced strain effect is discussed in more detail below.

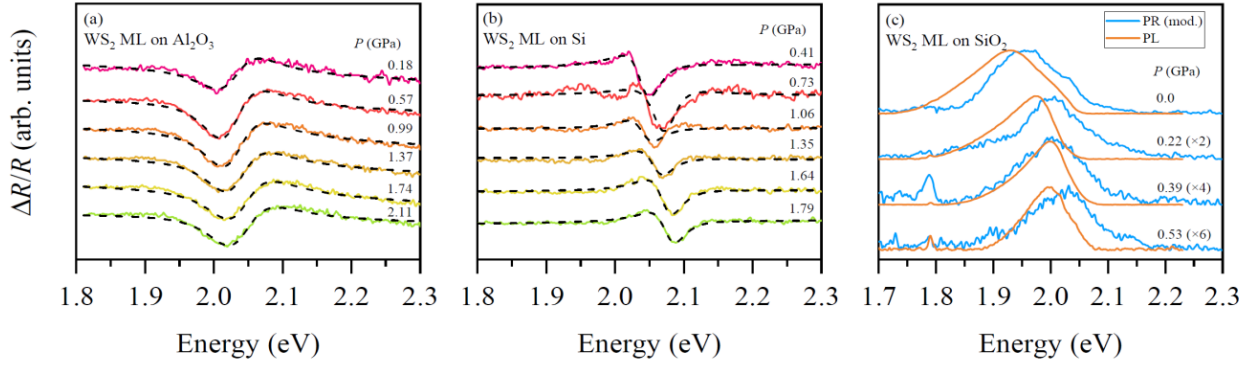

**Fig. 2. (a)** Photoreflectance spectra of a WS$_2$ monolayer deposited on sapphire acquired at different pressures. The energy of the transition A slightly increases with pressure. Black dashed curves are fits to the data. **(b)** Photoreflectance spectra of a WS$_2$ monolayer deposited on a silicon substrate acquired at different pressures. It can be seen that the energy of the transition A increases with pressure. Black dashed curves are fits to the data. **(c)** Moduli of the photoreflectance spectra of a WS$_2$ monolayer deposited on silicon (blue) and its corresponding photoluminescence (orange) emission. It can be seen that the energy of the A transition significantly increases with pressure (spectra at high pressure has been scaled several factors for better comparison).

Similarly, to the bulk case (see previous section), the PR spectra obtained for WS$_2$ monolayers at different pressures has been fitted with the Aspnes formula and the energy of the A transition was extracted for each pressure value. The pressure dependence of the A transition is plotted in Fig. 3 for samples deposited on Si (green symbols), sapphire (orange symbols) and glass (red symbols). Moreover, the energy of the PL peak has been included when PL signal could be measured (cross symbols). As it can be seen in Fig. 3, the pressure dependence of the A transition as extracted from PR measurements is very similar to that of PL, which is expected due to a small stokes shift, as previously discussed. From these values, linear fits were performed and pressure coefficients were extracted. These range from 11 meV/GPa for the sample deposited on sapphire up to 123 meV/GPa for the sample deposited on glass. The large dispersion in pressure coefficients is accounted for by the substrate-induced strain onto the layers. In this regard, it is physically more meaningful to evaluate the evolution of the band gap as a function of in-plane strain.

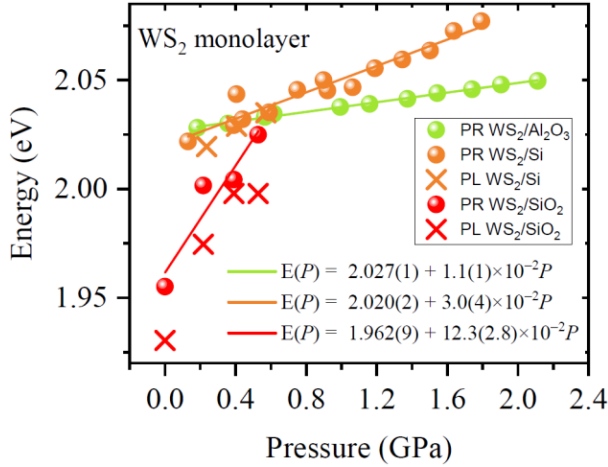

**Fig. 3.**

Pressure dependence of the fitted energy spectra of the A excitonic transition in monolayer $WS_2$ from the photoreflectance (PR) and photoluminescence (PL) spectra of samples deposited on sapphire (red), silicon (green) and glass (blue). Detected photoluminescence (PL) energies are shown as cross symbols. A linear fit to the data is used to extract the pressure coefficients. It can be seen that the pressure coefficient strongly depends on the substrate.

Here, we assume that all our layers are fully adhered to the substrate. This is to be expected for hydrostatic measurements since layer-substrate friction forces are several orders of magnitude higher than in bare tensile measurements. This is due to the fact that a strong normal force is present inside the hydrostatic chamber (the out-of-plane load per unit of area corresponds to the hydrostatic pressure $F_N = -P$).[71] Moreover, it is worth noting that the in-plane strain experienced by our samples is relatively small, around $\varepsilon = (a_{HP} - a)/a = -0.6\%$ (a being the lattice parameter and $a_{HP}$ the lattice parameter of $WS_2$ at the highest pressure), compared to typical fully adhered tensile measurements, which are free of out-of-plane forces (i.e. $F_N = 0$), around 1%.[7] More importantly, the assumption that the layers are fully adhered is confirmed by our first-principles calculations, which perfectly reproduce the experimental pressure coefficients when no slippage is considered (see discussion in following sections), i.e. no partial relaxation between the layer and the substrate takes place during the experiment.

Hence, assuming a total adhesion condition, the in-plane strain of the layers can be written as,

$$\varepsilon = \frac{-P}{3B}. \qquad [5]$$

where $B$ is the substrate's bulk modulus and $P$ is the pressure inside the high-pressure device.

Using Eq. [5] one can plot the pressure dependence of the band gap as a function of in-plane strain. This is shown in Fig. 4-a. As it can be seen in the figure, the gauge factor increases for those samples deposited on softer substrates, i.e. from 75 meV/% for the sample on sapphire up to 155 meV/% for the sample deposited on glass. These differences are accounted for by differences in out-of-plane forces. For instance, an in-plane strain of 0.3% corresponds to an out-of-plane force of ≈2 GPa for $WS_2/Al_2O_3$ and only of ≈1 GPa for WS$_2$/Si. Therefore, to fully understand the pressure dependence of the optical transitions in a high-pressure experiment, both components, in-plane and out-of-plane must be considered.

Here we propose a simple model to describe the pressure dependence of the band gap of multilayers on the bulk modulus of the substrate. Within the present model the variation of the band gap can be written as

$$\Delta E = \alpha \sigma_\perp + 2\beta \sigma_\parallel, \qquad [6]$$

where $\sigma_\perp$ and $\sigma_\parallel$ are out-of-plane and in-plane stress components, respectively, and $\alpha$ and $\beta$ are linear coefficients. For the case of a monolayer deposited on a substrate, $\sigma_\perp = P$, and the parallel component is related to the strain through the elastic tensor[72] as $\sigma_\parallel \approx (C_{11} + C_{12})\varepsilon_\parallel$, where $C_{11} = 233.4$ GPa and $C_{12} = 47.6$ GPa, as obtained from the calculated values of $C_{11}^{2D} = 144.88$ N/m and $C_{12}^{2D} = 29.53$ N/m (i.e. $C_{11} = C_{11}^{2D}/h$, $C_{12} = C_{12}^{2D}/h$, $h = 6.208$ Å). It should be noted that monolayer $C_{11}$ and $C_{12}$ are similar to our calculated values for bulk WS$_2$, 235.4 GPa and 50.4 GPa, respectively. Assuming that the sample's strain is the same as the substrate, it can be easily shown that the pressure coefficient can be written as

$$\frac{dE}{dP}(B) = \alpha + \frac{2\beta(C_{11} + C_{12})}{3B}. \qquad [7]$$

From this model it is clear that the pressure coefficient of layered systems deposited on substrates mostly depend on; *i*) two intrinsic parameters, $\alpha$ and $\beta$, which are related to the material's deformation potentials, *ii*) the elastic tensor components and *iii*) the bulk modulus of the substrate.

The pressure coefficient of our layers as a function of bulk modulus is plotted in Fig. 4-b), together with a data point obtained from the work of Han et al.[34] who measured the pressure coefficient of the PL signal of a WS$_2$ monolayer deposited on the hardest substrate, diamond. After fitting Eq. [6] to the experimental data (solid line in Fig. 4-b), we obtained $\alpha = -8(5)$ meV/GPa and $\beta = 24(2)$ meV/GPa. As it can be seen in the figure, the fit reproduces the experimental data points with great accuracy despite each data point corresponds to a substrate with a different dielectric constant (from 3.7 for SiO$_2$ up to 11.7 for Si). Hence the present model is valid regardless the substrate material. From our simplistic model, the pressure coefficient of a freestanding monolayer would be $dE/dP = \alpha + 2\beta$, which yields 40(6) meV/GPa, twice as large as the value reported for the bulk, consistent with our calculated theoretical value, around 42 meV/GPa (neglecting effects from the pressure transmitting medium), as presented below in the theoretical section.

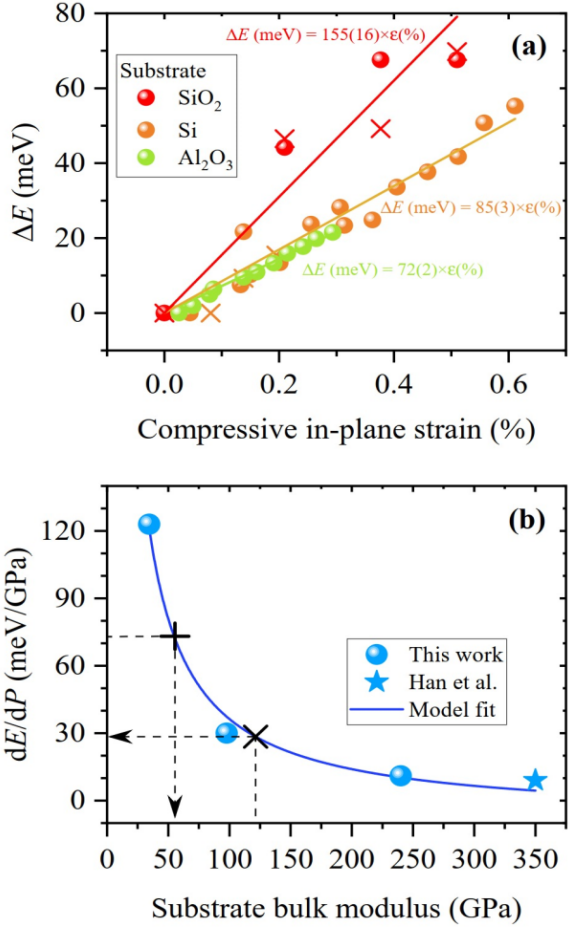

**Fig. 4.**

**(a)** Compressive strain dependence of the energy of the A excitonic transition of monolayer $WS_2$, as obtained from photoreflectance measurements for samples deposited on different substrates.

**(b)** Pressure coefficient of the A excitonic transition for monolayer $WS_2$ as a function of substrate bulk modulus (solid symbols) as obtained by means of PR in the present work and by means of PL elsewhere.[34] A proposed model (Eq. 6) is fitted to the data. Dashed lines indicate the pressure coefficient of freestanding monolayers and bulk.

Our model predicts a negative value for α, which results from two competing mechanisms that derive from the definition of the optical gap energy, $E_{opt} = E_{QP} - E_B$, where $E_{QP}$ is the quasiparticle (or band-to-band) transition energy and $E_B$ the excitonic binding energy. Upon out-of-plane stress, the quasiparticle gap contributes negatively to α while the excitonic binding energy diminishes with increasing stress, resulting in a positive contribution to α. More specifically, the excitonic binding energy can be roughly approximated using a Rydberg hydrogenic series; $E_b = 2\mu e^4/\hbar^2\varepsilon^2$,[73] where $\mu$ is the exciton reduced mass, and $\varepsilon$ is the effective dielectric constant that depends on the static dielectric constant of the environment, in this case, the PTM, Daphne 7474. As pressure increase, the volume of the Daphne 7474 strongly decrease resulting in an increased static dielectric constant, which effectively reduce $E_B$. Hence, it is expected that upon purely uniaxial compressive out-of-plane stress the optical band gap decreases. While many works studied the effect of in-plane stress on the band gap, to our knowledge no works have explored yet the effect of out-of-plane stress on the band gap of neither monolayer nor multilayered TMDCs.

For the purely in-plane biaxial strain case, it can be shown that the gauge factor is $dE/d\varepsilon_\parallel \approx 2\beta(C_{11}+C_{12})$, for the case of a $WS_2$ monolayer we found that $dE/d\varepsilon_\parallel = 135$

meV/%. This figure is in excellent agreement with theoretically predicted values, at 144 meV/%[54] and 130 meV/%,[74] as well as experimentally measured values in the tensile regime, −94 meV/%.[27] Similarly, under the uniaxial in-plane condition, $dE/d\varepsilon_\parallel \approx \beta(C_{11} + C_{12})$. In this scenario we find a $dE/d\varepsilon_\parallel = 56$ meV/% which is in good agreement with theoretically predicted values, 65 meV/%[74] as well as experimentally reported gauge factors measured under tensile conditions, −58.7 meV/%,[9] −45 meV/%[11] and −11 meV/%.[13] It is important to note that important discrepancies between theory and experiment might arise from slippage effects, which are important when soft substrates are used, where strain transfer rates as low as 12% have been reported under uniaxial tensile conditions,[12] as well as other sources of error such as funneling effects,[75] the presence of trions or a direct-to-indirect transition.[10,13] From our simple model, a factor of ≈2 is expected between the gauge factors of biaxial strain and uniaxial strain. This is in excellent agreement with theoretical calculations for $MoX_2$ and $WX_2$ (X=S, Se), as well as recent experiments, which found a factor of 2.3 for $MoS_2$ monolayers.[31,74] Finally, it is worth noting that owing to the hexagonal crystal structure both, stress[76] and deformation potentials[74] in TMDCs are quasi-isotropic in the in-plane directions (reported differences from armchair to zigzag directions are smaller than 2%).

### c) Multilayered $WS_2$ on sapphire

In order to evaluate the impact of friction forces on layered compounds, high-pressure measurements were performed on bilayers and trilayers. The PR spectra as acquired at different pressures is shown in Fig. 5 for a bilayer (top panel) and trilayer (bottom panel). As it can be seen in the figure, only one PR feature dominates the spectra, corresponding to the A exciton. The dashed black lines are Aspnes fits to the data, which allowed to extract the A energy at each pressure.

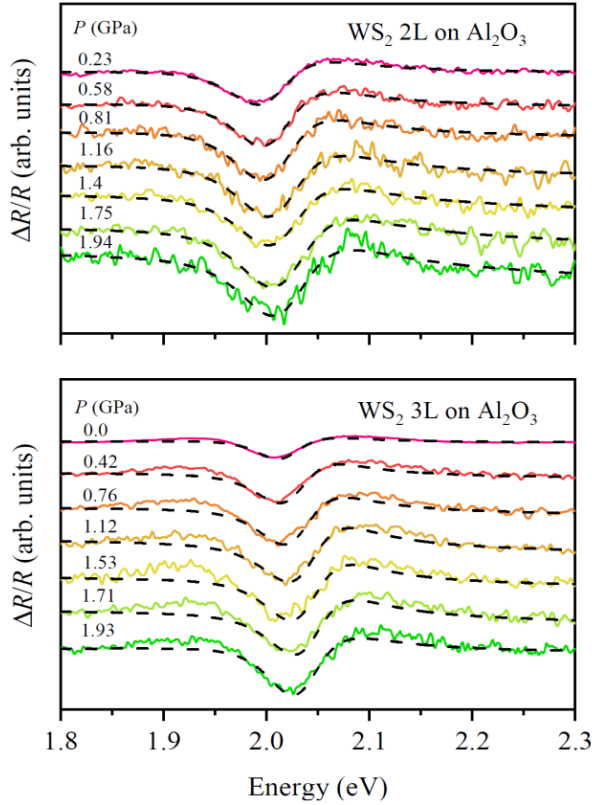

**Fig. 5.**

Photoreflectance spectra of a WS$_2$ bilayer (top panel) and trilayer (bottom panel) deposited on sapphire acquired at different pressures. The energy of the transition A slightly increases with pressure. Black dashed curves are fits to the data.

The pressure dependence of the A excitonic energy is plotted in Fig. 6-a) Strikingly it can be observed that the pressure coefficient is almost identical for monolayer, bilayer and trilayer deposited on sapphire. This result evidences that in-plane forces between the layers and the substrate are very strong and no partial relaxation takes place even for the trilayer case. For comparison purposes, Fig. 6 includes the pressure dependence of the A transition of bulk WS$_2$, which is significantly higher. The reduced pressure coefficient of the few-layered samples is due to the fact that these experience smaller in-plane strain as compared to the bulk freestanding case, due to the decreased compressibility of the substrate (i.e. sapphire) and their strong attachment to it. Indeed, when in-plane strain is considered instead of pressure (Fig. 6-b) the variation of energy is similar between layers on sapphire and bulk. For the bulk case, however, out-of-plane forces are smaller at a given strain value (relation to pressure was taken from ref.[77]), which results in a slightly larger slope than that of layered compounds on sapphire (note that this is mostly a consequence of a negative out-of-plane stress gauge, $\alpha < 0$).

While no partial slippage was observed in experimental conditions used in the present work, slippage is expected to take place under different conditions; *i)* the substrate exhibits a compressibility large enough and/or *ii)* samples exhibit a sufficiently large number of layers, which results in an increased rigidity of the multilayer. For the former condition, we estimate that the slippage regime can take place in WS$_2$ for substrates with bulk modulus $B < 34$ GPa (i.e. substrates more compressible than glass, already used in the present work). Previous high-pressure Raman measurements on graphene found partial relaxation for compressible substrates (i.e. SiO$_2$ and copper) while full adhesion took place for hard substrates (i.e. diamond and sapphire).[78] For the later condition, no evidence of slippage is found for a WS$_2$ trilayer deposited on sapphire, but it is expected to take place for a sufficiently high number of layers, $n$. For the case of graphite, partial relaxation was observed in samples with more than three layers on Si/SiO$_2$ substrates.[78] Hence, from our results it is clear that WS$_2$ exhibits a better adherence than graphene but the friction coefficient between the substrate and the sample is still unknown.

In the present work it was not possible to estimate the friction coefficient because no partial relaxation was measured. So far, tribology studies have been performed by means of atomic force microscopy (AFM) in two-dimensional materials such as TMDCs or graphene.[79–81] Here we propose that high-pressure methods could be likewise employed as a tool to determine the friction coefficient for different substrate compounds and number of layers (the friction coefficient has been shown to strongly depend on the number of layers[82,83,84]). To this end it would be highly desirable to perform high-pressure experiments of TMDCs in the slippage regime by either increasing the number of layers or using more compressible substrates.

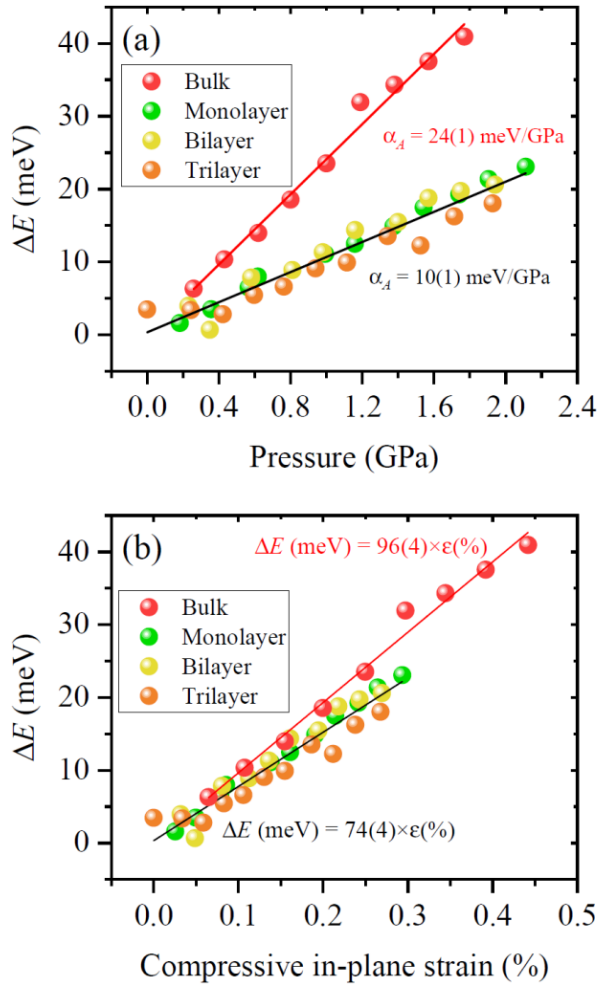

**Fig. 6.**

**(a)** Pressure dependence of the fitted energy of the A excitonic transition in a monolayer, bilayer and trilayer of $WS_2$ deposited on sapphire. The pressure coefficients are obtained by linear fits.

(b) Compressive in-plane strain dependence of the energy of the A excitonic transition of monolayer, bilayer and trilayer $WS_2$ deposited on sapphire.

The pressure coefficients of the direct transitions in $WS_2$ deposited on several substrates are shown in Table I, from monolayer to bulk. As it can be seen in the table, the pressure coefficients of $WS_2$ deposited on sapphire are very similar for the monolayer, bilayer, and trilayer case, around 10 meV/GPa, as previously discussed. The pressure coefficient of monolayers deposited on substrates with lower bulk modulus is much larger, up to 130 meV/GPa. The fact that the pressure coefficient is over an order of magnitude larger evidences that the substrate effects are critical in any high-pressure experiment for single- and multi-layers. The pressure coefficients of the A* feature as well as the B transition are included in the table for the bulk compound, and not shown for the monolayer due to lack of signal. For comparison purposes, Table I includes the calculated pressure coefficients simulating the strain conditions of the monolayers on different substrates (in parenthesis). It can be seen that, despite the large dispersion of measured pressure coefficients for samples on different substrates, the theoretical calculations reproduce with great accuracy the experimental figures. In this regard, it is important to note that disperse values are also found in the literature, where band gap and excitonic pressure coefficients, spanning from 30 (Ref. [34]) to 540 (Ref. [33]) meV/GPa, were typically provided without analyzing the substrate or dielectric environment effects as shown in Table S-III and S-IV in the Supplementary Material.

The relation between the experimental pressure coefficient, $dE_{opt}/dP$, with the DFT-calculated band-to-band transition $dE_{QP}/dP$ is given by,

$$\frac{dE_{opt}}{dP} = \frac{dE_{QP}}{dP} - \frac{dE_B}{dP} \qquad [8]$$

where $dE_B/dP$ is the pressure coefficient of the exciton binding energy. Our BSE calculations revealed that the excitonic contribution to the pressure coefficient of the optical gap is significant and strongly depends on two factors; *i)* the strain conditions (from 0 to 7.4 meV/GPa in this work) as well as *ii)* the dielectric environment (from −1.5 to −3.2 meV/GPa in this work). All these effects were considered for the calculated values shown in Table I.

| d$E$/d$P$ (meV/GPa) | Sapphire $B_0$ = 240 GPa | Silicon $B_0$ = 97.8 GPa | Glass $B_0$ = 34.5 GPa | Freestanding |
|---|---|---|---|---|
| **1 ML** | PR: 11±1 (15.3, 11.0[#]) | PR: 30±4 PL: 46±5 (44.7, 37.3[#]) | PR: 123±28 PL: 133±30 (145.6, 115.4[#]) | -- (73.1, 37[#]) |
| **2 L** | PR: 11±1 (13) | | | |
| **3 L** | PR: 8.5±1 (13) | | | |
| **Bulk** | | | | PR: 24±1 (33.7) PR: 36±3[#] (39.3*) PR: 34±2* (31.9[†], 42.2[††]) |

**Table I.** Photoreflectance (PR) and photoluminescence (PL) experimental values (calculated values are shown below, in parenthesis) for the pressure coefficient of the exciton A in WS$_2$, as well as for exciton B (marked with [#]) and exciton A* (marked with *). [†]For A($n$=2). [††]Calculated pressure coefficient of band-to-band transition in the H k-point of the BZ. All calculations include excitonic effects and dielectric screening effects from both the substrate and PTM (i.e. Daphne 7474), except for the B excitons where changes of ε($P$) are neglected in order to provide a general case for better reference. Bulk moduli, $B_0$, used in the present work are included for reference for each of the substrate material.

### d) DFT calculations

The calculated electronic band structure of bulk WS$_2$ at ambient and high pressure is shown in Fig. 7(a) along the high symmetry points of the Brillouin zone (shown on the panel c of the figure). As it can be seen in the figure, bulk WS$_2$ exhibits a fundamental indirect gap between the valence band maximum (VBM) at Γ and conduction band minimum (CBM) at a point in the Γ-K direction of the Brillouin zone. At K there are two direct excitonic transitions, namely A and B. Upon compression, we find that the indirect band gap decreases at a rate of −72.5 meV/GPa, in agreement with previous DFT calculations, which reported a negative pressure coefficient for this transition.[85] Another work showed that the pressure coefficient of the indirect transition was highly sensitive to the functional used, ranging from −26.2 meV/GPa up to −129 meV/GPa,[43] but the direct transitions blueshifts instead. Experimental measurements also found a direct-to-indirect transition for other TMDCs such as WSe$_2$ monolayer and bilayers, which exhibit a negative pressure coefficient of −3 and −22 meV/GPa, respectively,[36] or such as MoS$_2$, whose indirect band gap was estimated to decrease at a rate of −15.3 meV/GPa from PL measurements for pressures higher than the crossover at 1.9 GPa.[86] It is worth noting that bulk, TMDCs exhibit hidden spin-polarized bands even in their bulk, centrosymmetric form which were difficult to energetically resolve at ambient conditions,[3] however, our calculations predict a splitting of the conduction band minima at K at high pressure for WS$_2$, which can be exploited to study the physics behind dark excitons. Hence, the calculated pressure coefficients of the A transition reported in the present manuscript correspond to the transition at the K point between VBM and the second conduction band state, right above CBM.

The electronic band structure of 1L WS$_2$ calculated at ambient and high pressure is shown in Fig. 7-b. As it can be appreciated in the figure, the direct transition A at the K point is ≈300 meV below the indirect transition in the Γ-K direction. Our calculations and experiments show that at ambient pressure WS$_2$ is a direct semiconductor with PL excitonic emission. At higher pressure, the PL signal is quenched and vanishes at 0.73 GPa for a monolayer deposited on silicon, indicating a direct-to-indirect transition. This value is close to our calculated transition pressure, at 0.6 GPa for a freestanding layer. Similar transition pressures have been reported for other TMDCs such as WSe$_2$ or MoS$_2$.[36,86] It is worth noting that our calculations reveal that 1L, 2L and 3L remains indirect at 2 GPa regardless the choice of substrate (see fig S3 and S4 in the Supplementary Material).

From Table I, our calculations reveal that the pressure coefficient of a freestanding monolayer is around 73.1 meV/GPa (effects from Daphne PTM are included). This corresponds to the pressure coefficient that of a monolayer deposited on a substrate with a bulk modulus of 55.5 GPa (from Eq. [7]), as shown in Fig. 4-b (vertical cross). Hence, we propose to use substrates with bulk modulus around 55.5 GPa for future high-pressure experiments on monolayer and few-layer compounds. Such experiments would yield optical pressure coefficients similar to that of the freestanding case. Moreover, Similar strain conditions between the freestanding case and the deposited layer case are met when the substrate's bulk modulus is around 121 GPa. Hence we propose to use substrates with specific bulk modulus in order to simulate the optical and structural conditions of the freestanding monolayers as a natural alternative to experiments performed on solution-suspended layers which typically exhibit decreased optical properties due to the presence of defects or the solidification of the PTM. More interestingly, the in-plane compressibility of both bulk and multilayered $WS_2$ corresponds to the compressibility of a material with a bulk modulus of 121.4 GPa.[87] When one evaluates the pressure coefficient of a monolayer deposited on such substrate (from Eq. [7]), it is expected that its pressure coefficient is of 28 meV/GPa, very close to the experimentally measured pressure coefficient in bulk $WS_2$, at 24 meV/GPa despite our model is to be used only for two-dimensional systems. This can be visually seen in Fig. 4-b (diagonal cross). Such sticking similarity might indicate that excitonic effects (much more important for monolayers when compared to the bulk counterpart) do not strongly impact the pressure coefficient of the optical gap, and that the optical gap variations in TMDCs are mostly governed by the interatomic distances.

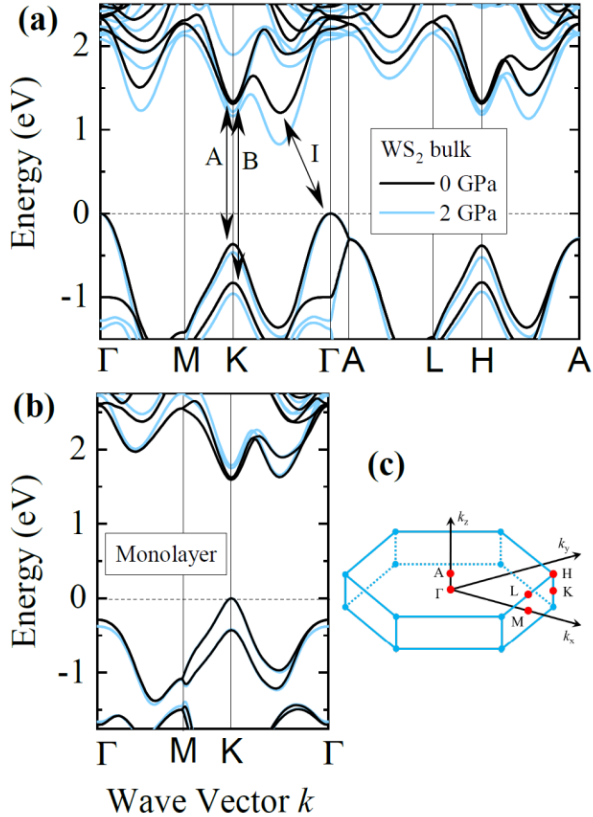

**Fig. 7.**

**a)** Electronic band structure of bulk $WS_2$ calculated along the main high-symmetry points of the Brillouin zone at a pressure of 0 GPa (black curves) and 2 GPa (blue curves). The A and B excitonic transitions are shown with arrows, as well as the indirect transition I.

**b)** Electronic band structure of a $WS_2$ monolayer calculated along the main high-symmetry points of the Brillouin zone at a pressure of 0 GPa (black curves) and 2 GPa (blue curves).

**c)** Schematic representation of the Brillouin zone of 2H-$WS_2$ and the positions of high-symmetry k-points.

### e) BSE calculations

The bulk excitonic binding energies and their pressure coefficients were obtained via the effective BSE as well as within Gerlach–Pollmann model for the 1s exciton state of the A and B transitions (shown in Table S-V of the SM). We found an excellent agreement for the A 1s state between both methods which validates the Gerlach–Pollmann model (GP) as a useful tool due to the low computational cost as compared to the BSE. However, the GP model can only be applied to the ground state excitons localized at the energy minima, which limits its applicability to vdW structures. Our calculations also show that the excitonic energies and pressure coefficients of the: i) A* interlayer exciton, ii) A 2s state and iii) band-to-band transition at the H point are similar (see Table I and Table S-VI in the SM) to the experimentally determined value of the A* feature, around 34(2) meV/GPa. This suggests that a combination of these three contributions could be responsible for the experimental signal. In order to shed new light into this issue, the in-plane and out-of-plane dispersions (from DFT) are displayed together with excitonic wave functions in reciprocal space (from BSE) and presented in Fig. S5 of the SM. These calculations allowed us to conclude that the exciton wave function of the A exciton is strongly localized around K, while the B exciton exhibits a large dispersion in the out-of-plane direction (i.e. along the K-H direction). Hence, the BSE calculations and robust GW-BSE calculations discard the presence of optically-active excitons localized around the H point, in agreement with low-temperature micro-reflectance contrast spectroscopy under high-magnetic fields,[68] where A* features of the $WX_2$ family were all assigned to interlayer transitions.

For the case of 1L $WS_2$, the excitonic binding energy is strongly influenced by the dielectric environment. As shown in Fig. 8-(a),(e) the calculated excitonic binding energy decreases with increasing dielectric constant of the surroundings of the monolayer, which in a typical high-pressure experiment corresponds to the PTM ($\varepsilon_{PTM}$) and the substrate ($\varepsilon_{subs}$), respectively. Since the excitonic binding energy is also parametrized by the effective mass, it is necessary to calculate its value for a given strain state of the monolayer, which in a high pressure experiment, is fully determined by the compressibility of the substrate (in the present work we assume full adhesion, as previously discussed). It is necessary to consider both effects, the dielectric environment and the effective mass in order to properly calculate the optical pressure coefficient. This can be clearly observed in Fig. 8 (b-d, f-h), where the pressure coefficient of the excitonic binding energies are calculated as a function of the dielectric constant of the PTM for different substrates, with (solid line) and without (dashed lines) effective mass corrections.

From Fig. 8, it can be seen that $dE_B/dP$ strongly depends on the dielectric constant of the PTM. For the case of a typical PTM, methanol-ethanol-water (16:3:1) mixture with a large static dielectric constant ($\varepsilon_{PTM} \approx 34$) the optical pressure coefficient can be approximated as that of band-to-band (i.e. $dE_{opt}/dP \approx dE_{QP}/dP$, from eq. [8]) with a calculated discrepancy lower than 2 meV/GPa for the here considered substrates. However, excitonic corrections $dE_B/dP$, become very important (up to 8 meV/GPa) for other typical PTMs with small dielectric constants such as noble gases (from He to Xe, in the range $\varepsilon_{PTM}$ = 1.06 – 1.88) or Daphne 7474 ($\varepsilon_{PTM} \approx 2.0$).[88] In the present work, the increase of the dielectric constant due to the pressure was also included in the calculations (for Daphne $\varepsilon_{PTM} \approx 2.6$ at 2 GPa, as estimated from its compressibility). Finally, it is worth noting that while only the A exciton was measured, the present analysis also holds for the B exciton (see Fig. 8(e-h)). Our calculations suggest that the excitonic effects in the pressure coefficient of the B exciton are even smaller than that of the A exciton, due to a larger reduced effective mass (see Tab. S-II in SM). From these calculations it can be concluded that the pressure coefficient of the optical transitions in high-pressure experiments is mostly influenced by substrate-induced strain and hydrostatic pressure rather than variations in the dielectric environment by approximately an order of magnitude, however the latter effect should not be neglected when using PTM with small dielectric constants.

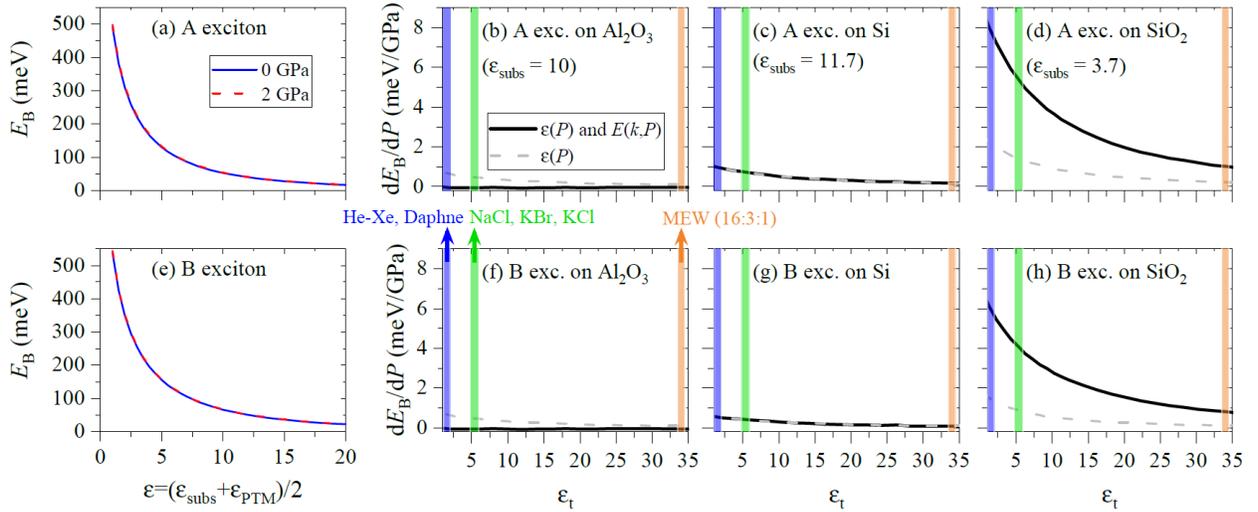

**Fig. 8. (a, e)** Exciton binding energies for A and B excitons in 1L WS$_2$ as function of the effective dielectric environment at pressures of 0 and 2 GPa (blue and red curves, respectively) for a freestanding monolayer. **(b-e and f-h)** Pressure coefficients for A and B excitons on sapphire, silicon and glass substrates as a function of the dielectric constant of the top material. Solid lines ("$\varepsilon(P)$ and $E(k,P)$") incorporate the electronic dispersion (effective mass) and dielectric properties of the substrates while dashed lines ("$\varepsilon(P)$") consider the substrate influence only in the dielectric confinement. The shaded regions indicates the dielectric constants of some typical pressure transmitting media.

## IV. Conclusions

In conclusion, our results show that few-layered structures remain fully adhered to the substrate even for incommensurate systems upon compression conditions. Hence, it is crucial to consider the substrate effect for any high-pressure experiment on two-dimensional materials. Indeed, our high-pressure optical experiments combined with first principles and effective BSE calculations show that multilayered $WS_2$ remains fully adhered at least up to a −0.6% in-plane compressive strain and a thickness ranging from monolayer to at least trilayer. The substrate-induced strain effect on the structural and optical properties can be larger than an order of magnitude. In the present work, we provide a simple physical model to describe the layer-substrate interaction upon stress condition. The present model allowed us to estimate compressive gauge factors for the in-plane and out of plane components, in the biaxial (135 meV/%) and uniaxial (56 meV/%) cases. More interestingly, we found that the effect of compressive pressure in the out-of-plane component results in an overall shrinking of the excitonic gap. Taking in consideration the effect of substrate strain in a high-pressure experiment, we estimate that the direct-to-indirect transition of a freestanding monolayer takes place at 0.53 GPa. The evolution of a closely-lying excitonic transition close to the A exciton, $A^*$, as well as the B exciton, is also discussed. Finally, our effective BSE calculations show that excitonic effects must be taken into consideration in order to accurately determine the pressure coefficient of strongly excitonic systems such as $WS_2$. In particular, different strain conditions and dielectric environments such as different substrates and pressure transmitting media can contribute at least up to ≈30% of the optical pressure coefficient.

**Supporting information**: Moduli of photoreflectance and photoluminescence spectra of a $WS_2$ monolayer acquired at different pressures. Band structures of bulk as well as 1L, 2L and 3L. Calculated A exciton and B exciton band structure in bulk $WS_2$. Tables of; band effectives masses and static dielectric tensor components, reduced mass for A and B excitons and screening length, previously reported pressure coefficients of excitonic transitions for selected bulk, mono- and multilayers TMDCs, calculated binding energies and pressure coefficients for A and B, 1s and 2s excitons, calculated pressure coefficients of excitons in bulk as well as monolayers on different substrates.

### Acknowledgements

This work was supported by the National Science Centre (NCN) Poland OPUS 11 no. 2016/21/B/ST3/00482. R.O. acknowledges the support by POLONEZ 3 no. 2016/23/P/ST3/04278. This project is carried out under POLONEZ program which has received funding from the European Union's Horizon 2020 research and innovation program under the Marie Sklodowska-Curie grant agreement No 665778. P.E.F.J. and J.F. acknowledge the financial support of the Deutsche Forschungsgemeinschaft (DFG, German Research Foundation) SFB 1277 (Project-ID 314695032, projects B07 and B11),

SPP 2244 (Project No. 443416183), and of the European Union Horizon 2020 Research and Innovation Program under Contract No. 881603 (Graphene Flagship). DFT calculations were carried out with the support of the Interdisciplinary Centre for Mathematical and Computational Modelling (ICM) University of Warsaw and Center for Information Services and High Performance Computing (ZIH) at TU Dresden.

# References

(1) Kolobov, A. V.; Tominaga, J. *Two-Dimensional Transition-Metal Dichalcogenides*; Springer Series in Materials Science; Springer International Publishing, 2016.

(2) Zhang, X.; Liu, Q.; Luo, J.-W.; Freeman, A. J.; Zunger, A. Hidden Spin Polarization in Inversion-Symmetric Bulk Crystals. *Nat. Phys.* **2014**, *10* (5), 387–393. https://doi.org/10.1038/nphys2933.

(3) Oliva, R.; Wozniak, T.; Dybala, F.; Kopaczek, J.; Scharoch, P.; Kudrawiec, R. Hidden Spin-Polarized Bands in Semiconducting 2H-MoTe$_2$. *Mater. Res. Lett.* **2020**, *8*, 75–81. https://doi.org/10.1080/21663831.2019.1702113.

(4) Akinwande, D.; Petrone, N.; Hone, J. Two-Dimensional Flexible Nanoelectronics. *Nat. Commun.* **2014**, *5* (1), 5678. https://doi.org/10.1038/ncomms6678.

(5) Jariwala, D.; Sangwan, V. K.; Lauhon, L. J.; Marks, T. J.; Hersam, M. C. Emerging Device Applications for Semiconducting Two-Dimensional Transition Metal Dichalcogenides. *ACS Nano* **2014**, *8* (2), 1102–1120. https://doi.org/10.1021/nn500064s.

(6) Schmidt, R.; Niehues, I.; Schneider, R.; Drüppel, M.; Deilmann, T.; Rohlfing, M.; Vasconcellos, S. M. de; Castellanos-Gomez, A.; Bratschitsch, R. Reversible Uniaxial Strain Tuning in Atomically Thin WSe$_2$. *2D Mater.* **2016**, *3* (2), 021011. https://doi.org/10.1088/2053-1583/3/2/021011.

(7) Island, J. O.; Kuc, A.; Diependaal, E. H.; Bratschitsch, R.; Zant, H. S. J. van der; Heine, T.; Castellanos-Gomez, A. Precise and Reversible Band Gap Tuning in Single-Layer MoSe$_2$ by Uniaxial Strain. *Nanoscale* **2016**, *8* (5), 2589–2593. https://doi.org/10.1039/C5NR08219F.

(8) Yang, R.; Lee, J.; Ghosh, S.; Tang, H.; Sankaran, R. M.; Zorman, C. A.; Feng, P. X.-L. Tuning Optical Signatures of Single- and Few-Layer MoS$_2$ by Blown-Bubble Bulge Straining up to Fracture. *Nano Lett.* **2017**, *17* (8), 4568–4575. https://doi.org/10.1021/acs.nanolett.7b00730.

(9) Wang, F.; Li, S.; Bissett, M. A.; Kinloch, I. A.; Li, Z.; Young, R. J. Strain Engineering in Monolayer WS$_2$ and WS$_2$ Nanocomposites. *2D Mater.* **2020**, *7* (4), 045022. https://doi.org/10.1088/2053-1583/ababf1.

(10) Niehues, I.; Marauhn, P.; Deilmann, T.; Wigger, D.; Schmidt, R.; Arora, A.; Vasconcellos, S. M. de; Rohlfing, M.; Bratschitsch, R. Strain Tuning of the Stokes Shift in Atomically Thin Semiconductors. *Nanoscale* **2020**, *12* (40), 20786–20796. https://doi.org/10.1039/D0NR04557H.

(11) He, X.; Li, H.; Zhu, Z.; Dai, Z.; Yang, Y.; Yang, P.; Zhang, Q.; Li, P.; Schwingenschlogl, U.; Zhang, X. Strain Engineering in Monolayer WS$_2$, MoS$_2$, and the WS$_2$/MoS$_2$ Heterostructure. *Appl. Phys. Lett.* **2016**, *109* (17), 173105. https://doi.org/10.1063/1.4966218.

(12) Zhang, Q.; Chang, Z.; Xu, G.; Wang, Z.; Zhang, Y.; Xu, Z.-Q.; Chen, S.; Bao, Q.; Liu, J. Z.; Mai, Y.-W.; Duan, W.; Fuhrer, M. S.; Zheng, C. Strain Relaxation of Monolayer WS$_2$ on Plastic Substrate. *Adv. Funct. Mater.* **2016**, *26* (47), 8707–8714. https://doi.org/10.1002/adfm.201603064.

(13) Wang, Y.; Cong, C.; Yang, W.; Shang, J.; Peimyoo, N.; Chen, Y.; Kang, J.; Wang, J.; Huang, W.; Yu, T. Strain-Induced Direct–Indirect Bandgap Transition and Phonon Modulation in Monolayer WS$_2$. *Nano Res.* **2015**, *8* (8), 2562–2572. https://doi.org/10.1007/s12274-015-0762-6.

(14) Liu, Z.; Amani, M.; Najmaei, S.; Xu, Q.; Zou, X.; Zhou, W.; Yu, T.; Qiu, C.; Birdwell, A. G.; Crowne, F. J.; Vajtai, R.; Yakobson, B. I.; Xia, Z.; Dubey, M.; Ajayan, P. M.; Lou, J. Strain and Structure Heterogeneity in MoS2 Atomic Layers Grown by Chemical Vapour Deposition. *Nat. Commun.* **2014**, *5*, 5246. https://doi.org/10.1038/ncomms6246.

(15) Conley, H. J.; Wang, B.; Ziegler, J. I.; Haglund, R. F.; Pantelides, S. T.; Bolotin, K. I. Bandgap Engineering of Strained Monolayer and Bilayer MoS$_2$. *Nano Lett.* **2013**, *13* (8), 3626–3630. https://doi.org/10.1021/nl4014748.

(16) He, K.; Poole, C.; Mak, K. F.; Shan, J. Experimental Demonstration of Continuous Electronic Structure Tuning via Strain in Atomically Thin MoS$_2$. *Nano Lett.* **2013**, *13* (6), 2931–2936. https://doi.org/10.1021/nl4013166.

(17) Zhu, C. R.; Wang, G.; Liu, B. L.; Marie, X.; Qiao, X. F.; Zhang, X.; Wu, X. X.; Fan, H.; Tan, P. H.; Amand, T.; Urbaszek, B. Strain Tuning of Optical Emission Energy and Polarization in Monolayer and Bilayer MoS$_2$. *Phys. Rev. B* **2013**, *88* (12), 121301. https://doi.org/10.1103/PhysRevB.88.121301.

(18) Wang, Y.; Cong, C.; Qiu, C.; Yu, T. Raman Spectroscopy Study of Lattice Vibration and Crystallographic Orientation of Monolayer MoS$_2$ under Uniaxial Strain. *Small* **2013**, *9* (17), 2857–2861. https://doi.org/10.1002/smll.201202876.

(19) Rice, C.; Young, R. J.; Zan, R.; Bangert, U.; Wolverson, D.; Georgiou, T.; Jalil, R.; Novoselov, K. S. Raman-Scattering Measurements and First-Principles Calculations of Strain-Induced Phonon Shifts in Monolayer MoS$_2$. *Phys. Rev. B* **2013**, *87* (8), 081307. https://doi.org/10.1103/PhysRevB.87.081307.

(20) Blundo, E.; Felici, M.; Yildirim, T.; Pettinari, G.; Tedeschi, D.; Miriametro, A.; Liu, B.; Ma, W.; Lu, Y.; Polimeni, A. Evidence of the Direct-to-Indirect Band Gap Transition in Strained Two-Dimensional WS$_2$, MoS$_2$, and WSe$_2$. *Phys. Rev. Res.* **2020**, *2* (1), 012024. https://doi.org/10.1103/PhysRevResearch.2.012024.

(21) Guo, Y.; Li, B.; Huang, Y.; Du, S.; Sun, C.; Luo, H.; Liu, B.; Zhou, X.; Yang, J.; Li, J.; Gu, C. Direct Bandgap Engineering with Local Biaxial Strain in Few-Layer MoS$_2$ Bubbles. *Nano Res.* **2020**, *13* (8), 2072–2078. https://doi.org/10.1007/s12274-020-2809-6.

(22) Tedeschi, D.; Blundo, E.; Felici, M.; Pettinari, G.; Liu, B.; Yildrim, T.; Petroni, E.; Zhang, C.; Zhu, Y.; Sennato, S.; Lu, Y.; Polimeni, A. Controlled Micro/Nanodome Formation in Proton-Irradiated Bulk Transition-Metal Dichalcogenides. *Adv. Mater.* **2019**, *31* (44), 1903795. https://doi.org/10.1002/adma.201903795.

(23) Tyurnina, A. V.; Bandurin, D. A.; Khestanova, E.; Kravets, V. G.; Koperski, M.; Guinea, F.; Grigorenko, A. N.; Geim, A. K.; Grigorieva, I. V. Strained Bubbles in van Der Waals Heterostructures as Local Emitters of Photoluminescence with Adjustable Wavelength. *ACS Photonics* **2019**, *6* (2), 516–524. https://doi.org/10.1021/acsphotonics.8b01497.

(24) Ryu, Y. K.; Carrascoso, F.; López-Nebreda, R.; Agraït, N.; Frisenda, R.; Castellanos-Gomez, A. Microheater Actuators as a Versatile Platform for Strain Engineering in 2D Materials. *Nano Lett.* **2020**, *20* (7), 5339–5345. https://doi.org/10.1021/acs.nanolett.0c01706.

(25) Carrascoso, F.; Lin, D.-Y.; Frisenda, R.; Castellanos-Gomez, A. Biaxial Strain Tuning of Interlayer Excitons in Bilayer $MoS_2$. *J. Phys. Mater.* **2019**, *3* (1), 015003. https://doi.org/10.1088/2515-7639/ab4432.

(26) Gant, P.; Huang, P.; Pérez de Lara, D.; Guo, D.; Frisenda, R.; Castellanos-Gomez, A. A Strain Tunable Single-Layer $MoS_2$ Photodetector. *Mater. Today* **2019**, *27*, 8–13. https://doi.org/10.1016/j.mattod.2019.04.019.

(27) Frisenda, R.; Drüppel, M.; Schmidt, R.; Michaelis de Vasconcellos, S.; Perez de Lara, D.; Bratschitsch, R.; Rohlfing, M.; Castellanos-Gomez, A. Biaxial Strain Tuning of the Optical Properties of Single-Layer Transition Metal Dichalcogenides. *Npj 2D Mater. Appl.* **2017**, *1* (1), 1–7. https://doi.org/10.1038/s41699-017-0013-7.

(28) Plechinger, G.; Castellanos-Gomez, A.; Buscema, M.; Zant, H. S. J. van der; Steele, G. A.; Kuc, A.; Heine, T.; Schüller, C.; Korn, T. Control of Biaxial Strain in Single-Layer Molybdenite Using Local Thermal Expansion of the Substrate. *2D Mater.* **2015**, *2* (1), 015006. https://doi.org/10.1088/2053-1583/2/1/015006.

(29) Hui, Y. Y.; Liu, X.; Jie, W.; Chan, N. Y.; Hao, J.; Hsu, Y.-T.; Li, L.-J.; Guo, W.; Lau, S. P. Exceptional Tunability of Band Energy in a Compressively Strained Trilayer $MoS_2$ Sheet. *ACS Nano* **2013**, *7* (8), 7126–7131. https://doi.org/10.1021/nn4024834.

(30) Li, H.; Contryman, A. W.; Qian, X.; Ardakani, S. M.; Gong, Y.; Wang, X.; Weisse, J. M.; Lee, C. H.; Zhao, J.; Ajayan, P. M.; Li, J.; Manoharan, H. C.; Zheng, X. Optoelectronic Crystal of Artificial Atoms in Strain-Textured Molybdenum Disulphide. *Nat. Commun.* **2015**, *6* (1), 7381. https://doi.org/10.1038/ncomms8381.

(31) Carrascoso, F.; Frisenda, R.; Castellanos-Gomez, A. Biaxial versus Uniaxial Strain Tuning of Single-Layer $MoS_2$. *Nano Mater. Sci.* **2021**. https://doi.org/10.1016/j.nanoms.2021.03.001.

(32) Michail, A.; Anestopoulos, D.; Delikoukos, N.; Parthenios, J.; Grammatikopoulos, S.; Tsirkas, S. A.; Lathiotakis, N. N.; Frank, O.; Filintoglou, K.; Papagelis, K. Biaxial Strain Engineering of CVD and Exfoliated Single- and Bi-Layer $MoS_2$ Crystals. *2D Mater.* **2020**, *8* (1), 015023. https://doi.org/10.1088/2053-1583/abc2de.

(33) Kim, J.-S.; Ahmad, R.; Pandey, T.; Rai, A.; Feng, S.; Yang, J.; Lin, Z.; Mauricio Terrones; Banerjee, S. K.; Singh, A. K.; Akinwande, D.; Lin, J.-F. Towards Band Structure and Band Offset Engineering of Monolayer Mo $(1- x)$ W $(x)$ S 2 via Strain. *2D Mater.* **2018**, *5* (1), 015008. https://doi.org/10.1088/2053-1583/aa8e71.

(34) Han, B.; Li, F.; Li, L.; Huang, X.; Gong, Y.; Fu, X.; Gao, H.; Zhou, Q.; Cui, T. Correlatively Dependent Lattice and Electronic Structural Evolutions in Compressed Monolayer Tungsten Disulfide. *J. Phys. Chem. Lett.* **2017**, *8* (5), 941–947. https://doi.org/10.1021/acs.jpclett.7b00133.

(35) Shen, P.; Ma, X.; Guan, Z.; Li, Q.; Huafang, Z.; Liu, R.; Liu, B.; Yang, X.; Dong, Q.; Cui, T.; Liu, B. Linear Tunability of the Band Gap and Two-Dimensional (2D) to Three-Dimensional (3D) Isostructural Transition in WSe2 under High Pressure. *J Phys Chem C* **2017**, *121*, 26019–26026.

(36) Ye, Y.; Dou, X.; Ding, K.; Jiang, D.; Yang, F.; Sun, B. Pressure-Induced K–Λ Crossing in Monolayer WSe2. *Nanoscale* **2016**, *8* (20), 10843–10848. https://doi.org/10.1039/C6NR02690G.

(37) Li, F.; Yan, Y.; Han, B.; Li, L.; Huang, X.; Yao, M.; Gong, Y.; Jin, X.; Liu, B.; Zhu, C.; Zhou, Q.; Cui, T. Pressure Confinement Effect in MoS2 Monolayers. *Nanoscale* **2015**, *7* (19), 9075–9082. https://doi.org/10.1039/C5NR00580A.

(38) Francisco-López, A.; Han, B.; Lagarde, D.; Marie, X.; Urbaszek, B.; Robert, C.; Goñi, A. On the Impact of the Stress Situation on the Optical Properties of WSe$_2$ Monolayers under High Pressure. *Pap. Phys.* **2019**, *11*, 110005–110005. https://doi.org/10.4279/pip.110005.

(39) Suski, T.; Paul, W. *High Pressure in Semiconductor Physics I and II, Semiconductors and Semimetals*; Academic Press, 1998; Vol. Vols. 54 and 55.

(40) Zelewski, S. J.; Kudrawiec, R. Photoacoustic and Modulated Reflectance Studies of Indirect and Direct Band Gap in van Der Waals Crystals. *Sci. Rep.* **2017**, *7* (1), 15365. https://doi.org/10.1038/s41598-017-15763-1.

(41) Kopaczek, J.; Polak, M. P.; Scharoch, P.; Wu, K.; Chen, B.; Tongay, S.; Kudrawiec, R. Direct Optical Transitions at K- and H-Point of Brillouin Zone in Bulk MoS2, MoSe2, WS2, and WSe2. *J. Appl. Phys.* **2016**, *119* (23), 235705. https://doi.org/10.1063/1.4954157.

(42) Kopaczek, J.; Zelewski, S. J.; Polak, M. P.; Gawlik, A.; Chiappe, D.; Schulze, A.; Caymax, M.; Kudrawiec, R. Direct and Indirect Optical Transitions in Bulk and Atomically Thin MoS2 Studied by Photoreflectance and Photoacoustic Spectroscopy. *J. Appl. Phys.* **2019**, *125* (13), 135701. https://doi.org/10.1063/1.5080300.

(43) Dybała, F.; Polak, M. P.; Kopaczek, J.; Scharoch, P.; Wu, K.; Tongay, S.; Kudrawiec, R. Pressure Coefficients for Direct Optical Transitions in MoS2, MoSe2, WS2, and WSe2 Crystals and Semiconductor to Metal Transitions. *Sci. Rep.* **2016**, *6*, 26663. https://doi.org/10.1038/srep26663.

(44) Teutenberg, T.; Wiese, S.; Wagner, P.; Gmehling, J. High-Temperature Liquid Chromatography. Part III: Determination of the Static Permittivities of Pure Solvents and Binary Solvent Mixtures—Implications for Liquid Chromatographic Separations. *J. Chromatogr. A* **2009**, *1216* (48), 8480–8487. https://doi.org/10.1016/j.chroma.2009.09.076.

(45) Kudrawiec, R.; Misiewicz, J. Photoreflectance Spectroscopy of Semiconductor Structures at Hydrostatic Pressure: A Comparison of GaInAs/GaAs and GaInNAs/GaAs Single Quantum Wells. *Appl. Surf. Sci.* **2006**, *253* (1), 80–84. https://doi.org/10.1016/j.apsusc.2006.05.073.

(46) Kresse, G.; Joubert, D. From Ultrasoft Pseudopotentials to the Projector Augmented-Wave Method. *Phys. Rev. B* **1999**, *59* (3), 1758–1775. https://doi.org/10.1103/PhysRevB.59.1758.

(47) Kresse, G.; Furthmüller, J. Efficiency of Ab-Initio Total Energy Calculations for Metals and Semiconductors Using a Plane-Wave Basis Set. *Comput. Mater. Sci.* **1996**, *6* (1), 15–50. https://doi.org/10.1016/0927-0256(96)00008-0.

(48) Perdew, J. P.; Burke, K.; Ernzerhof, M. Generalized Gradient Approximation Made Simple. *Phys. Rev. Lett.* **1996**, *77* (18), 3865–3868. https://doi.org/10.1103/PhysRevLett.77.3865.

(49) Grimme, S.; Antony, J.; Ehrlich, S.; Krieg, H. A Consistent and Accurate Ab Initio Parametrization of Density Functional Dispersion Correction (DFT-D) for the 94 Elements H-Pu. *J. Chem. Phys.* **2010**, *132* (15), 154104. https://doi.org/10.1063/1.3382344.

(50) Choudhary, K.; Zhang, Q.; Reid, A. C. E.; Chowdhury, S.; Van Nguyen, N.; Trautt, Z.; Newrock, M. W.; Congo, F. Y.; Tavazza, F. Computational Screening of High-Performance Optoelectronic Materials Using OptB88vdW and TB-MBJ Formalisms. *Sci. Data* **2018**, *5*, 180082. https://doi.org/10.1038/sdata.2018.82.

(51) Rohlfing, M.; Louie, S. G. Electron-Hole Excitations in Semiconductors and Insulators. *Phys. Rev. Lett.* **1998**, *81* (11), 2312–2315. https://doi.org/10.1103/PhysRevLett.81.2312.

(52) Rohlfing, M.; Louie, S. G. Electron-Hole Excitations and Optical Spectra from First Principles. *Phys. Rev. B* **2000**, *62* (8), 4927–4944. https://doi.org/10.1103/PhysRevB.62.4927.

(53) Faria Junior, P. E.; Kurpas, M.; Gmitra, M.; Fabian, J. Kp Theory for Phosphorene: Effective g-Factors, Landau Levels, and Excitons. *Phys. Rev. B* **2019**, *100* (11), 115203. https://doi.org/10.1103/PhysRevB.100.115203.

(54) Zollner, K.; Junior, P. E.; Fabian, J. Strain-Tunable Orbital, Spin-Orbit, and Optical Properties of Monolayer Transition-Metal Dichalcogenides. *Phys. Rev. B* **2019**, *100* (19), 195126. https://doi.org/10.1103/PhysRevB.100.195126.

(55) Birowska, M.; Faria Junior, P. E.; Fabian, J.; Kunstmann, J. Large Exciton Binding Energies in $MnPS_3$ as a Case Study of a van Der Waals Layered Magnet. *Phys. Rev. B* **2021**, *103* (12), L121108. https://doi.org/10.1103/PhysRevB.103.L121108.

(56) Landau, L. D.; Pitaevskii, L. P.; Lifshitz, E. M.; Bell, J.; Kearsley, M.; Sykes, J. *Electrodynamics of Continuous Media*; Elsevier, 2013; Vol. 8.

(57) Tedeschi, D.; De Luca, M.; Faria Junior, P. E.; Granados del Águila, A.; Gao, Q.; Tan, H. H.; Scharf, B.; Christianen, P. C. M.; Jagadish, C.; Fabian, J.; Polimeni, A. Unusual Spin Properties of InP Wurtzite Nanowires Revealed by Zeeman Splitting Spectroscopy. *Phys. Rev. B* **2019**, *99* (16), 161204. https://doi.org/10.1103/PhysRevB.99.161204.

(58) Rytova, N. S. The Screened Potential of a Point Charge in a Thin Film. *Mosc. Univ. Phys. Bull.* **1967**, *3*, 18.

(59) Keldysh, L. V. Coulomb Interaction in Thin Semiconductor and Semimetal Films. *Sov. J. Exp. Theor. Phys. Lett.* **1979**, *29*, 658.

(60) Gerlach, B.; Pollmann, J. Binding Energies and Wave Functions of Wannier Excitons in Uniaxial Crystals — a Modified Perturbation Approach. I. Theory. *Phys. Status Solidi B* **1975**, *67* (1), 93–103. https://doi.org/10.1002/pssb.2220670107.

(61) Aspnes, D. E. Third-Derivative Modulation Spectroscopy with Low-Field Electroreflectance. *Surf. Sci.* **1973**, *37* (Supplement C), 418–442. https://doi.org/10.1016/0039-6028(73)90337-3.

(62) Molas, M. R.; Nogajewski, K.; Slobodeniuk, A. O.; Binder, J.; Bartos, M.; Potemski, M. The Optical Response of Monolayer, Few-Layer and Bulk Tungsten Disulfide. *Nanoscale* **2017**, *9* (35), 13128–13141. https://doi.org/10.1039/C7NR04672C.

(63) Arora, A.; Drüppel, M.; Schmidt, R.; Deilmann, T.; Schneider, R.; Molas, M. R.; Marauhn, P.; Vasconcellos, S. M. de; Potemski, M.; Rohlfing, M.; Bratschitsch, R. Interlayer Excitons in a Bulk van Der Waals Semiconductor. *Nat. Commun.* **2017**, *8* (1), 639. https://doi.org/10.1038/s41467-017-00691-5.

(64) Jindal, V.; Bhuyan, S.; Deilmann, T.; Ghosh, S. Anomalous Behavior of the Excited State of the A Exciton in Bulk $WS_2$. *Phys. Rev. B* **2018**, *97* (4), 045211. https://doi.org/10.1103/PhysRevB.97.045211.

(65) Lin, K.-I.; Chen, Y.-J.; Wang, B.-Y.; Cheng, Y.-C.; Chen, C.-H. Photoreflectance Study of the Near-Band-Edge Transitions of Chemical Vapor Deposition-Grown Mono- and Few-Layer $MoS_2$ Films. *J. Appl. Phys.* **2016**, *119* (11), 115703. https://doi.org/10.1063/1.4944437.

(66) Saigal, N.; Ghosh, S. H-Point Exciton Transitions in Bulk MoS2. *Appl. Phys. Lett.* **2015**, *106* (18), 182103. https://doi.org/10.1063/1.4920986.

(67) Zhang, C.; Wang, H.; Chan, W.; Manolatou, C.; Rana, F. Absorption of Light by Excitons and Trions in Monolayers of Metal Dichalcogenide $MoS_2$: Experiments and Theory. *Phys. Rev. B* **2014**, *89* (20), 205436. https://doi.org/10.1103/PhysRevB.89.205436.

(68) Arora, A.; Deilmann, T.; Marauhn, P.; Drüppel, M.; Schneider, R.; Molas, M. R.; Vaclavkova, D.; Michaelis de Vasconcellos, S.; Rohlfing, M.; Potemski, M.; Bratschitsch, R. Valley-Contrasting Optics of Interlayer Excitons in Mo- and W-Based Bulk Transition Metal Dichalcogenides. *Nanoscale* **2018**, *10* (33), 15571–15577. https://doi.org/10.1039/c8nr03764g.

(69) Brotons-Gisbert, M.; Segura, A.; Robles, R.; Canadell, E.; Ordejón, P.; Sánchez-Royo, J. F. Optical and Electronic Properties of 2H-$MoS_2$ under Pressure: Revealing the Spin-Polarized Nature of Bulk Electronic Bands. *Phys. Rev. Mater.* **2018**, *2* (5), 054602. https://doi.org/10.1103/PhysRevMaterials.2.054602.

(70) Niehues, I.; Schmidt, R.; Drüppel, M.; Marauhn, P.; Christiansen, D.; Selig, M.; Berghäuser, G.; Wigger, D.; Schneider, R.; Braasch, L.; Koch, R.; Castellanos-Gomez, A.; Kuhn, T.; Knorr, A.; Malic, E.; Rohlfing, M.; Michaelis de

Vasconcellos, S.; Bratschitsch, R. Strain Control of Exciton-Phonon Coupling in Atomically Thin Semiconductors. *Nano Lett.* **2018**, *18* (3), 1751–1757. https://doi.org/10.1021/acs.nanolett.7b04868.

(71) Peng, J.; Yang, F.; Huang, K.; Dong, H.; Yan, S.; Zheng, X. Friction Behavior of Monolayer Molybdenum Diselenide Nanosheet under Normal Electric Field. *Phys. Lett. A* **2020**, *384* (7), 126166. https://doi.org/10.1016/j.physleta.2019.126166.

(72) Zeng, F.; Zhang, W.-B.; Tang, B.-Y. Electronic Structures and Elastic Properties of Monolayer and Bilayer Transition Metal Dichalcogenides $MX_2$ ( M = Mo, W; X = O, S, Se, Te): A Comparative First-Principles Study. *Chin. Phys. B* **2015**, *24* (9), 097103. https://doi.org/10.1088/1674-1056/24/9/097103.

(73) Chernikov, A.; Berkelbach, T. C.; Hill, H. M.; Rigosi, A.; Li, Y.; Aslan, O. B.; Reichman, D. R.; Hybertsen, M. S.; Heinz, T. F. Exciton Binding Energy and Nonhydrogenic Rydberg Series in Monolayer $WS_2$. *Phys. Rev. Lett.* **2014**, *113* (7), 076802. https://doi.org/10.1103/PhysRevLett.113.076802.

(74) Song, W.; Yang, L. Quasiparticle Band Gaps and Optical Spectra of Strained Monolayer Transition-Metal Dichalcogenides. *Phys. Rev. B* **2017**, *96* (23), 235441. https://doi.org/10.1103/PhysRevB.96.235441.

(75) Chaves, A.; Azadani, J. G.; Alsalman, H.; da Costa, D. R.; Frisenda, R.; Chaves, A. J.; Song, S. H.; Kim, Y. D.; He, D.; Zhou, J.; Castellanos-Gomez, A.; Peeters, F. M.; Liu, Z.; Hinkle, C. L.; Oh, S.-H.; Ye, P. D.; Koester, S. J.; Lee, Y. H.; Avouris, P.; Wang, X.; Low, T. Bandgap Engineering of Two-Dimensional Semiconductor Materials. *Npj 2D Mater. Appl.* **2020**, *4* (1), 1–21. https://doi.org/10.1038/s41699-020-00162-4.

(76) Li, T. Ideal Strength and Phonon Instability in Single-Layer $MoS_2$. *Phys. Rev. B* **2012**, *85* (23), 235407. https://doi.org/10.1103/PhysRevB.85.235407.

(77) Selvi, E.; Ma, Y.; Aksoy, R.; Ertas, A.; White, A. High Pressure X-Ray Diffraction Study of Tungsten Disulfide. *J. Phys. Chem. Solids* **2006**, *67* (9), 2183–2186. https://doi.org/10.1016/j.jpcs.2006.05.008.

(78) Machon, D.; Bousige, C.; Alencar, R.; Torres-Dias, A.; Balima, F.; Nicolle, J.; Pinheiro, G. de S.; Filho, A. G. S.; San-Miguel, A. Raman Scattering Studies of Graphene under High Pressure. *J. Raman Spectrosc.* **2018**, *49* (1), 121–129. https://doi.org/10.1002/jrs.5284.

(79) Andersson, D.; de Wijn, A. S. Understanding the Friction of Atomically Thin Layered Materials. *Nat. Commun.* **2020**, *11* (1), 420. https://doi.org/10.1038/s41467-019-14239-2.

(80) Sattari Baboukani, B.; Ye, Z.; G. Reyes, K.; Nalam, P. C. Prediction of Nanoscale Friction for Two-Dimensional Materials Using a Machine Learning Approach. *Tribol. Lett.* **2020**, *68* (2), 57. https://doi.org/10.1007/s11249-020-01294-w.

(81) Akinwande, D.; Brennan, C. J.; Bunch, J. S.; Egberts, P.; Felts, J. R.; Gao, H.; Huang, R.; Kim, J.-S.; Li, T.; Li, Y.; Liechti, K. M.; Lu, N.; Park, H. S.; Reed, E. J.; Wang, P.; Yakobson, B. I.; Zhang, T.; Zhang, Y.-W.; Zhou, Y.; Zhu, Y. A Review on Mechanics and Mechanical Properties of 2D Materials—Graphene and Beyond. *Extreme Mech. Lett.* **2017**, *13*, 42–77. https://doi.org/10.1016/j.eml.2017.01.008.

(82) Fang, L.; Liu, D.-M.; Guo, Y.; Liao, Z.-M.; Luo, J.-B.; Wen, S.-Z. Thickness Dependent Friction on Few-Layer $MoS_2$, $WS_2$, and $WSe_2$. *Nanotechnology* **2017**, *28* (24), 245703. https://doi.org/10.1088/1361-6528/aa712b.

(83) Lee, C.; Li, Q.; Kalb, W.; Liu, X.-Z.; Berger, H.; Carpick, R. W.; Hone, J. Frictional Characteristics of Atomically Thin Sheets. *Science* **2010**, *328* (5974), 76–80. https://doi.org/10.1126/science.1184167.

(84) Filleter, T.; McChesney, J. L.; Bostwick, A.; Rotenberg, E.; Emtsev, K. V.; Seyller, Th.; Horn, K.; Bennewitz, R. Friction and Dissipation in Epitaxial Graphene Films. *Phys. Rev. Lett.* **2009**, *102* (8), 086102. https://doi.org/10.1103/PhysRevLett.102.086102.

(85) Shang, J.; Chen, P.; Zhang, L.; Fengxiao, Z.; Xuerui, C. The Electronic and Optical Properties of Tungsten Disulfide under High Pressure. *Chem. Phys. Lett.* **2016**, *651*, 257–260.

(86) Fu, L.; Wan, Y.; Tang, N.; Ding, Y.; Gao, J.; Yu, J.; Guan, H.; Zhang, K.; Wang, W.; Zhang, C.; Shi, J.; Wu, X.; Shi, S.-F.; Ge, W.; Dai, L.; Shen, B. K-Λ Crossover Transition in the Conduction Band of Monolayer $MoS_2$ under Hydrostatic Pressure. *Sci. Adv.* **2017**, *3* (11), e1700162. https://doi.org/10.1126/sciadv.1700162.

(87) Li, L.; Zeng, Z.; Liang, T.; Tang, M.; Cheng, Y. Elastic Properties and Electronic Structure of $WS_2$ under Pressure from First-Principles Calculations. **2016**. https://doi.org/10.1515/zna-2016-0398.

(88) Amey, R. L.; Cole, R. H. Dielectric Constants of Liquefied Noble Gases and Methane. *J. Chem. Phys.* **1964**, *40* (1), 146–148. https://doi.org/10.1063/1.1724850.

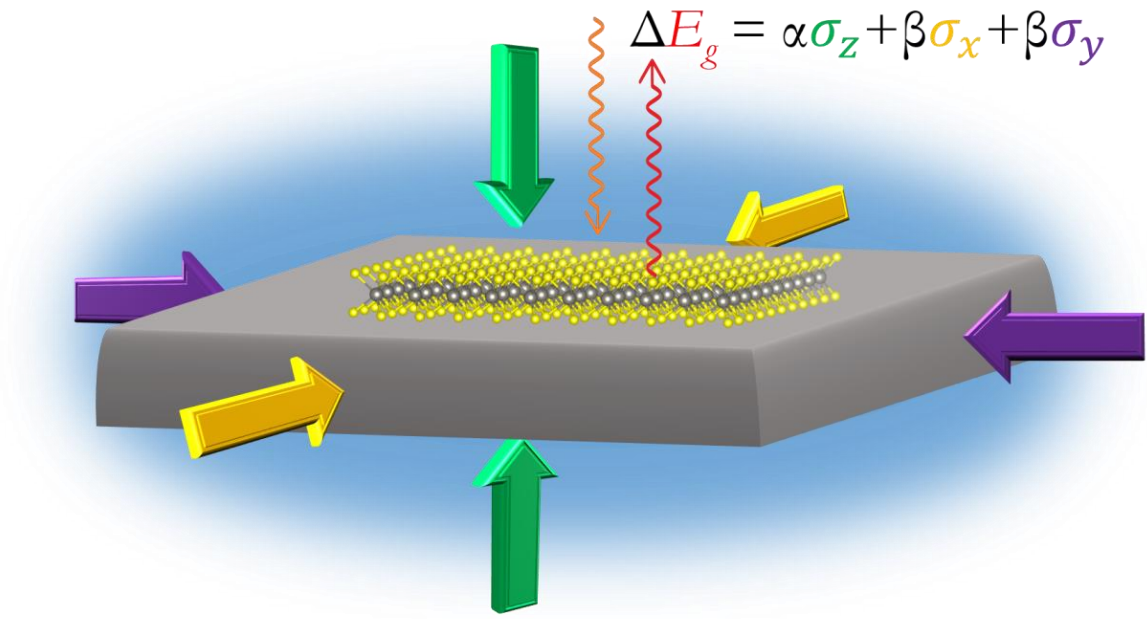

**Graphic of Manuscript**

# Supporting Information: Strong substrate strain effects in multilayered WS₂ revealed by high-pressure optical measurements

Robert Oliva,[1*] Tomasz Wozniak,[1] Paulo Eduardo de Faria Junior,[2] Filip Dybala,[1] Jan Kopaczek,[1] Jaroslav Fabian,[2] Paweł Scharoch,[1] Robert Kudrawiec[1]

[1] Department of Semiconductor Materials Engineering, Faculty of Fundamental Problems of Technology, Wroclaw University of Science and Technology, Wybrzeże Wyspiańskiego 27, 50-370 Wrocław, Poland

[2] Department of Physics, University of Regensburg, 93040 Regensburg, Germany.

[*] Corresponding author: robert.oliva.vidal@pwr.edu.pl

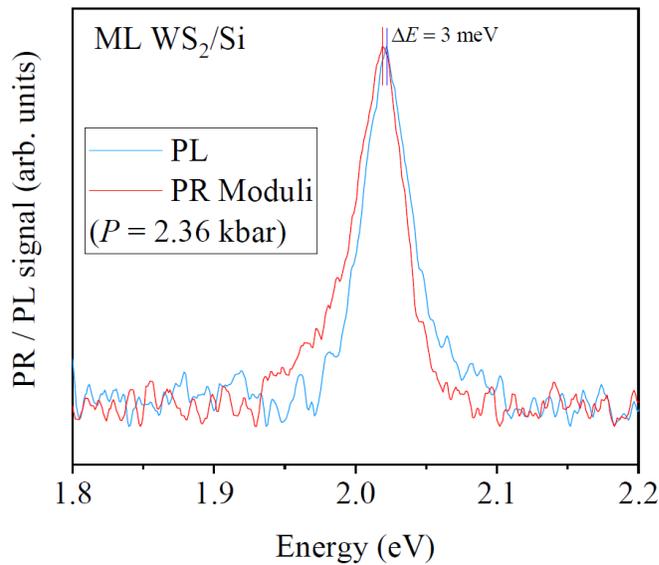

**Fig. S1.**

Moduli of photoreflectance and photoluminescence spectra of a WS₂ monolayer deposited on a Si substrate acquired at a pressure of 0.236 GPa. A very small Stokes shift (≈3 meV) can be observed.

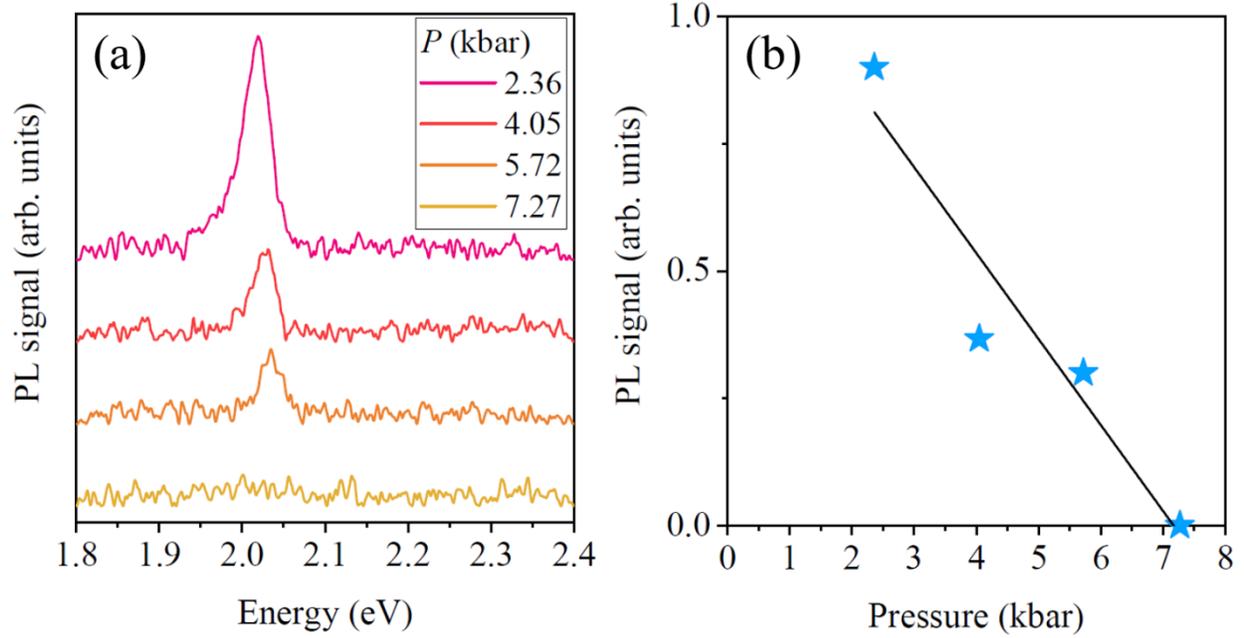

**Fig. S2.** (a) Photoluminescence spectra of a WS$_2$ monolayer deposited on Si acquired at different pressures, a smooth substrate has been subtracted for clarity. (b) The PL signal monotonically decrease with increasing pressure, vanishing at 0.72 GPa.

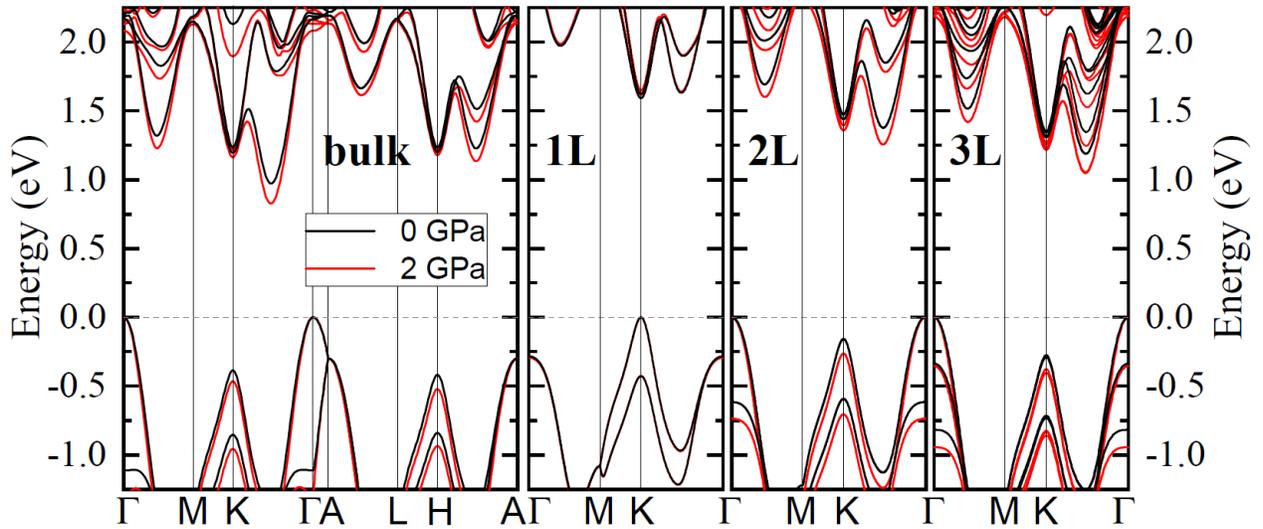

**Fig. S3.** Band structures of bulk as well as 1L, 2L and 3L WS$_2$ on a sapphire substrate at pressures of 0 and 2 GPa.

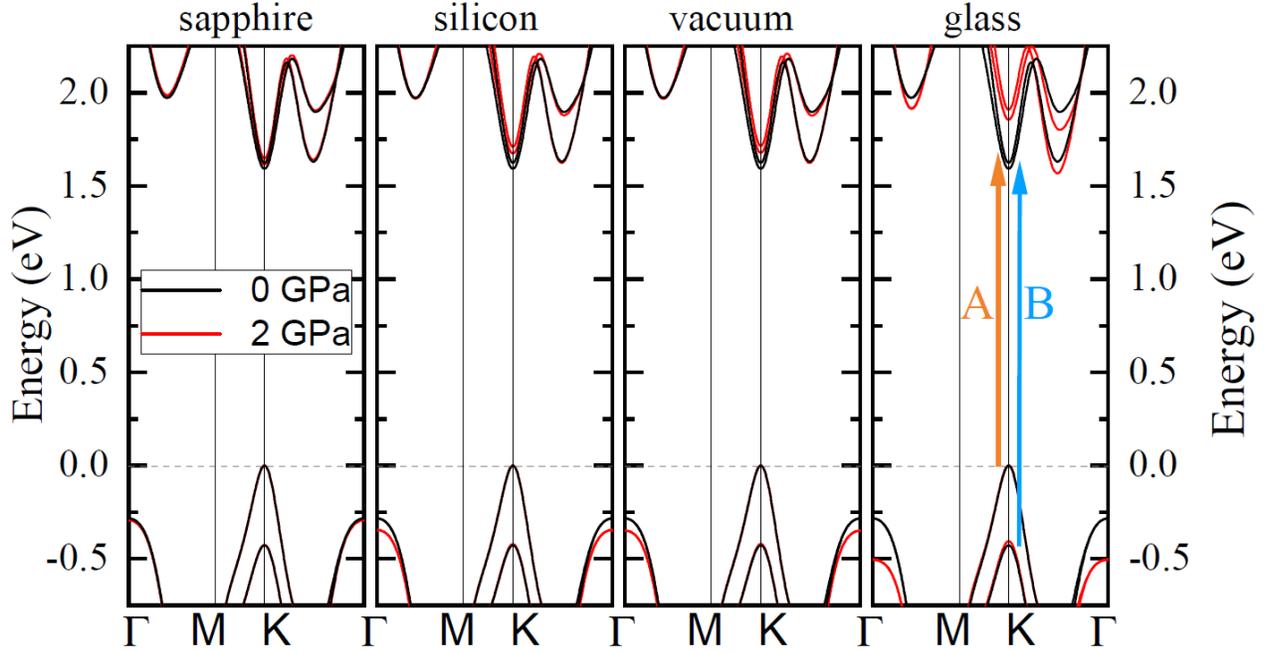

**Fig. S4.** Band structures of 1L WS$_2$ on different substrates at pressures of 0 and 2 GPa.

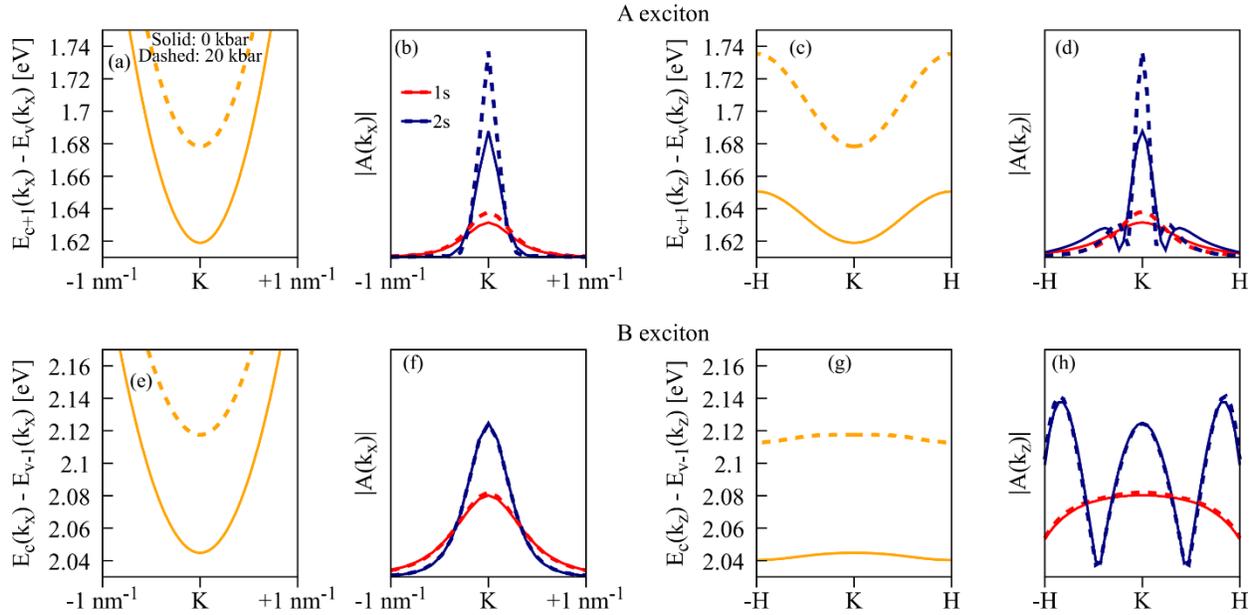

**Fig. S5. Top panel**: A exciton. **Bottom panel**: B exciton in bulk WS$_2$. **(a) and (e)**: Energy difference of conduction and valence bands in $k_Z=0$ plane and **(c), (g)** along the $k_Z$ direction. **(b) and (f)** Exciton wave functions of 1s (red) and 2s (blue) states in $k_Z=0$ plane and **(d), (h)** along $k_Z$ direction. 0 and 20 kbar pressures are represented by solid and dashed lines, respectively.

| P (kbar) | $m_{v-1}$ | $m_v$ | $m_c$ | $m_{c+1}$ | $\varepsilon_{xx}$ | $\varepsilon_{zz}$ |
|---|---|---|---|---|---|---|
| 0 | -0.48 | -0.35 | 0.36 | 0.27 | 14.084 | 5.987 |
| 20 | -0.48 | -0.36 | 0.37 | 0.28 | 14.470 | 6.879 |

**Table S-I.** Band effectives masses and static dielectric tensor components calculated for bulk $WS_2$ at 0 and 20 kbar.

| | | 1L | | | 2L | | 3L | |
|---|---|---|---|---|---|---|---|---|
| substrate | P (kbar) | $\mu_A$ | $\mu_B$ | $\varrho_0$ | $\mu_A$ | $\mu_B$ | $\mu_A$ | $\mu_B$ |
| freestanding | 0 | 0.150 | 0.207 | 39.74 | 0.151 | 0.208 | 0.152 | 0.209 |
| freestanding | 20 | 0.154 | 0.209 | 39.32 | | | | |
| sapphire | 20 | 0.149 | 0.206 | 39.55 | 0.152 | 0.207 | 0.153 | 0.209 |
| silicon | 20 | 0.154 | 0.209 | 39.34 | | | | |
| glass | 20 | 0.168 | 0.222 | 38.89 | | | | |

**Table S-II.** Calculated parameters for 1L, 2L and 3L $WS_2$ at 0 and 20 kbar, in which $\mu_A$ is the reduced mass for the X=A, B exciton and $\varrho_0$ is the screening length (2D polarizability).

| Reference | dE/dP (meV/GPa) | E (eV) | Comments |
|---|---|---|---|
| Dybała et al.[1] | 27 | 1.993 | PR, transition A. |
| Dybała et al.[1] | 41 | 2.432 | PR, transition B. |
| Dybała et al.[1] | -61.1 | 1.12 (approx.) | Calc, Indirect (?-G) |
| Shang et al.[2] | −28.4 | | Calc. indirect gap. (M?-G) |
| Shang et al.[2] | −16.5 | | Calc. direct gap at M (trans A) (structure 2Hc) |
| Shang et al.[2] | 23 | | Calc. direct gap at H |
| Shang et al.[2] | −20 | | Calc. direct gap at G |
| Shen et al[3] | −24 | | $WSe_2$, absorbance |

**Table S-III.** Reported pressure coefficients of excitonic transitions for selected bulk TMDCs.

| Reference | dE/dP (meV/GPa) | E (eV) | Comments |
|---|---|---|---|
| Han et al. [4] | 0.9 | | Trans. A. On Diamond. Abs. |
| Han et al. [4] | 0.7 | | Trans. B. On Diamond. Abs. |
| Han et al. [4] | 2.1-2.25 | | Trans A. On Si/SiO2, PL |
| Han et al. [4] | 0.53-1.24 | | Trans A. On Diamond, PL |
| Han et al. [4] | 1.04 | | Trion. On Si/SiO2, PL |
| Han et al. [4] | 0.3 | | Trion. On Diamond. PL |
| Han et al. | 0.54 | | Trans. A. On Diamond. Abs. |
| Kim et al. [5] | 3.35-5.4 | 1.946 | PL, on Si/SiO2 |
| Han et al. [4] | 0.31 cm$^{-1}$/GPa | | A1', on Diamond |
| Han et al. [4] | 0.39 cm$^{-1}$/GPa | | A1', on Si/SiO2 |
| Ye et al. [6] | 3.15 | | WSe$_2$. PL. Trans at K. |
| Ye et al. [6] | 2.7 | | WSe$_2$, Bilayer. PL. Trans at K. |
| Ye et al. [6] | −0.3 | | WSe$_2$, Indirect Λ–K, monolayer |
| Ye et al. [6] | −2.2 | | WSe$_2$, Indirect Λ–K, bilayer |

**Table S-IV.** Reported pressure coefficients of excitonic transitions for mono- and multilayers of WS$_2$ (unless otherwise stated in the comment column) deposited on different substrates.

| | E$_B$ at 0 kbar (meV) | E$_B$ at 20 kbar (meV) | dE$_B$/dP (meV/GPa) |
|---|---|---|---|
| A1s | 43.3 (41.2) | 35.4 (33.1) | -4 |
| A2s | 14.8 | 10.4 | -2.2 |
| B1s | 63.4 | 57.7 | -2.9 |
| B2s | 30.8 | 29.1 | -0.9 |

**Table S-V.** Binding energies for A and B, 1s and 2s, excitons and their pressure coefficients in bulk WS$_2$ calculated from BSE. Values in parentheses are obtained from Gerlach-Pollmann model.

|  | dE$_{QP}$/dP | dE$_B$/dP | dE/dP | dE/dP experiment |
|---|---|---|---|---|
| A | 29.7 | -4 | 33.7 | 24±1 |
| B | 36.4 | -2.9 | 39.3 | 36±3 |
| A* as A2s | 29.7 | -2.2 | 31.9 | 34±2 |
| A* as H-point | 42.2 |  |  |  |

**Table S-VI.** Calculated pressure coefficients (in meV/GPa) of A, B and A* excitons in bulk WS$_2$ with band-to-band (from DFT) and excitonic (from BSE) contributions. Experimental values with uncertainties are included.

|  | dE$_{QP}$/dP | dE$_B$/dP ε(P) and E(k,P) | dE$_B$/dP ε(P) | dE/dP | dE/dP experiment |
|---|---|---|---|---|---|
| freestanding | 45.5 | 3.8 | 3.8 | 41.7 |  |
| sapphire | 12.1 | -0.1 | 1.3 | 12.2 – 10.8 | 11±1 |
| silicon | 43.2 | 0.9 | 0.9 | 42.3 | 30±4 |
| glass | 142.5 | 7.4 | 2.0 | 135.1 – 140.5 | 123±28 |

**Table S-VII.** Calculated pressure coefficients (in meV/GPa) of A exciton in 1L WS$_2$ with band-to-band (from DFT) and excitonic (from BSE) contributions neglecting the contribution from the pressure-dependence of the dielectric constant of the pressure transmitting media. Experimental values with uncertainties are included.

|  | dE$_{QP}$/dP | dE$_B$/dP ε(P) and E(k,P) | dE$_B$/dP ε(P) | dE/dP | dE/dP experiment |
|---|---|---|---|---|---|
| freestanding | 45.5 | -27.58 |  | 73.1 |  |
| sapphire | 12.1 | -3.18 | -2.04 | 15.3 | 11±1 |
| silicon | 43.2 | -1.53 | -1.54 | 44.7 | 30±4 |
| glass | 142.5 | -3.07 | -8.29 | 145.6 | 123±28 |

**Table S-VIII.** Calculated pressure coefficients (in meV/GPa) of A exciton in 1L WS$_2$ with band-to-band (from DFT) and excitonic (from BSE) contributions including the contribution from the pressure-dependence of the dielectric constant of the pressure transmitting media, Daphne 7474. Experimental values with uncertainties are included.

**References**

1. Dybała, F. *et al.* Pressure coefficients for direct optical transitions in $MoS_2$, $MoSe_2$, $WS_2$, and $WSe_2$ crystals and semiconductor to metal transitions. *Sci. Rep.* **6**, 26663 (2016).

2. Shang, J., Chen, P., Zhang, L., Fengxiao, Z. & Xuerui, C. The electronic and optical properties of Tungsten Disulfide under high pressure. *Chem. Phys. Lett.* **651**, 257–260 (2016).

3. Shen, P. *et al.* Linear Tunability of the Band Gap and Two-Dimensional (2D) to Three-Dimensional (3D) Isostructural Transition in WSe2 under High Pressure. *J Phys Chem C* **121**, 26019–26026 (2017).

4. Han, B. *et al.* Correlatively Dependent Lattice and Electronic Structural Evolutions in Compressed Monolayer Tungsten Disulfide. *J. Phys. Chem. Lett.* **8**, 941–947 (2017).

5. Kim, J.-S. *et al.* Towards band structure and band offset engineering of monolayer Mo (1− x ) W ( x ) S 2 via Strain. *2D Mater.* **5**, 015008 (2018).

6. Ye, Y. *et al.* Pressure-induced K–Λ crossing in monolayer WSe2. *Nanoscale* **8**, 10843–10848 (2016).
7